 \documentclass[preprint2]{aastex6}
\usepackage{graphicx}
\usepackage{subfigure}
\usepackage[flushleft]{threeparttable}
\usepackage{arydshln}
\usepackage{multirow}
\usepackage{natbib}
\usepackage{amsmath,amsfonts,amsthm,bm}
\usepackage[flushleft]{threeparttable}
\usepackage{hyperref}
\newcommand\aastex{AAS\TeX}

\shorttitle{\aastex\ Quasar photo-$z$ regression and candidate selection}
\shortauthors{Yang et al.}
\begin{document}

\title{Quasar Photometric Redshifts and Candidate Selection: A New Algorithm Based on Optical and Mid-Infrared Photometric Data}


\author{Qian Yang\altaffilmark{1,2,3}, Xue-Bing Wu\altaffilmark{1,2}, Xiaohui Fan\altaffilmark{3,2}, Linhua Jiang\altaffilmark{2}, Ian McGreer\altaffilmark{3}, Richard Green\altaffilmark{3}, Jinyi Yang\altaffilmark{1,2}, Jan-Torge Schindler\altaffilmark{3}, Feige Wang\altaffilmark{1,2}, Wenwen Zuo\altaffilmark{4}, Yuming Fu\altaffilmark{1,2}}

\altaffiltext{1}{Department of Astronomy, School of Physics, Peking University, Beijing 100871, China}
\altaffiltext{2}{Kavli Institute for Astronomy and Astrophysics, Peking University, Beijing 100871, China}
\altaffiltext{3}{Steward Observatory, University of Arizona, 933 North Cherry Avenue, Tucson, AZ 85721, USA}
\altaffiltext{4}{Shanghai Astronomical Observatory, Chinese Academy of Sciences, Shanghai 200030, China}

\begin{abstract}
We present a new algorithm to estimate quasar photometric redshifts (photo-$z$s), by considering the asymmetries in the relative flux distributions of quasars. The relative flux models are built with multivariate Skew-t distributions in the multi-dimensional space of relative fluxes as a function of redshift and magnitude. For 151,392 quasars in the SDSS, we achieve a photo-$z$ accuracy, defined as the fraction of quasars with the difference between the photo-$z$ $z_p$ and the spectroscopic redshift $z_s$, $|\Delta z| = |z_s-z_p|/(1+z_s)$ within 0.1, of 74\%. Combining the WISE W1 and W2 infrared data with the SDSS data, the photo-$z$ accuracy is enhanced to 87\%. Using the Pan-STARRS1 or DECaLS photometry with WISE W1 and W2 data, the photo-$z$ accuracies are 79\% and 72\%, respectively. The prior probabilities as a function of magnitude for quasars, stars and galaxies are calculated respectively based on (1) the quasar luminosity function; (2) the Milky Way synthetic simulation with the Besan\c{c}on model; (3) the Bayesian Galaxy Photometric Redshift estimation. The relative fluxes of stars are obtained with the Padova isochrones, and the relative fluxes of galaxies are modeled through galaxy templates. We test our classification method to select quasars using the DECaLS $g$, $r$, $z$, and WISE W1 and W2 photometry. The quasar selection completeness is higher than 70\% for a wide redshift range $0.5<z<4.5$, and a wide magnitude range $18<r<21.5$ mag. Our photo-$z$ regression and classification method has the potential to extend to future surveys. The photo-$z$ code will be publicly available.
\end{abstract}

\keywords{catalogs --- cosmology: observations  --- galaxies: distances and redshifts --- methods: statistical --- quasars: general}

\section{Introduction} \label{sec:introduction}
Quasars are among the most powerful objects in the Universe, found from low redshift to redshifts beyond 7 \citep{Mortlock2011}. Tracing the properties of quasars can help understand supermassive black holes in massive galaxies and the coevolution of black holes and their host galaxies \citep{Kormendy2013}. Large quasar surveys are important for finding the clustering of quasars and lensed quasars, and for probing the galaxy merger scenario and measuring the mass distribution of halos \citep[e.g.,][]{Oguri2006, Hennawi2010}. So far, more than 346,000 quasars have been spectroscopically identified in the SDSS \citep{Schneider2010, Paris2017}.

Massive spectroscopic surveys require a large amount of telescope time, so it is usually very expensive to obtain spectroscopic redshifts for large quasar samples. Photometric redshifts (photo-$z$s), derived from photometric data, provide an alternative technique to measure redshifts. Photometric quasar samples have been used to do many important studies, such as the clustering of quasars \citep{Myers2006, Myers2007a, Myers2007b}, quasar number count statistics \citep[e.g.,][]{Richards2009b, Richards2015}, cosmic magnification \citep{Scranton2005}, and the Integrated Sachs-Wolfe effect \citep{Giannantonio2006}. Besides, photo-$z$ estimation is very useful for quasar candidate selection in spectroscopic redshift surveys \citep{Richards2004, Richards2009a, Richards2009b, Richards2015}.

Nowadays, more and more photometric data are being acquired. For example, the Pan-STARRS1 Telescope \citep[PS1;][]{Kaiser2002} carried out a distinct set of imaging synoptic sky surveys that are useful for quasar searches in the southern sky. In the near future, the Large Synoptic Survey Telescope \citep[LSST;][]{Tyson2002} will bring more opportunities for photo-$z$ estimates and cosmology research based on photo-$z$ quasar samples. The Dark Energy Spectroscopic Instrument \citep[DESI;][]{DESI1,DESI2} is the successor to the Stage-III BOSS redshift survey, and will study baryon acoustic oscillations (BAO) and the growth of structure through redshift-space distortions (RSD) with a wide-area galaxy and quasar redshift survey. High efficiency quasar candidate selection would save a lot of follow-up observation time. We aim to improve the photo-$z$ accuracy of quasars and develop an efficient quasar candidate selection algorithm for a wide range of redshift and magnitude. With carefully defined selection completeness and efficiency correction, a photometrically selected quasar sample has the potential to be used to derive the quasar luminosity function (QLF), and reach a fainter magnitude limit than a spectroscopically identified sample. Moreover, photometrically selected quasars combined with multi-epoch and multi-band LSST data will be powerful for studies such as measuring black hole mass through photometric reverberation mapping \citep[e.g.,][]{Hernitschek2015, Zu2016}; detecting changing-look quasars \citep[e.g.,][]{Gezari2017}; and characterizing the variability of quasars \citep[e.g.,][]{MacLeod2010, Zuo2012}.

Different methods have been put forward to estimate the photo-$z$s of quasars, including quasar template fitting \citep[e.g.,][]{Budavari2001, Babbedge2004, Salvato2009}, the empirical color-redshift relation (CZR) \citep[e.g.,][]{Richards2001a, Wu2004, Weinstein2004, Wu2010, Wu2012}, Machine Learning \citep[e.g.,][]{Ball2007, Yeche2010, Laurino2011, Brescia2013, Zhang2013} and the XDQSOz method \citep{Bovy2012}. In the COSMOS field, the template fitting method is efficient with the photometry from 30 bands. But there are few fields with such rich photometry available. Apart from the template fitting method, the photo-$z$ regression method needs a training sample, usually a spectroscopically identified quasar sample. The redshift and magnitude distributions of spectroscopically identified quasars are affected by their target selection methods and the incompleteness of spectroscopic observations. So, dividing the spectroscopically identified quasar training sample into a grid of redshift and magnitude is helpful, considering the dependence of quasar colors on redshift and luminosity. Quasars are usually bluer when brighter, and the equivalent width (EW) of their emission lines are anti-correlated with the continuum flux \citep[Baldwin effect;][]{Baldwin1997}. The slope of the power law continuum, the EW and FWHM of emission lines span wide ranges \citep[e.g.,][]{Vanden2001, Telfer2002}. In addition, the redward flux of the Lyman-$\alpha$ emission profile in a quasar spectrum is affected by the absorption lines of the Lyman-$\alpha$ forest from neutral hydrogen along the line-of-sight to the quasar. The color distribution of quasars, even in a narrow redshift and magnitude bin, differs from a Gaussian distribution. It is obviously skewed and shows tails even when excluding broad absorption line (BAL) quasars. A significant population of red quasars exists \citep[e.g.,][]{Webster1995, Richards2001b, Richards2003, Hopkins2004}. \citet{Richards2003} defined a quasar to be dust-reddened with relative color $\Delta (g^*-i^*)$ redder than 0.2, corresponding to $E(B-V)=0.04$, and find 6\% quasars fall into the redder quasar category. Dust reddening at the redshift of the quasar is the primary explanation for the red tail in quasar color distribution. \citet{Hopkins2004} modeled the color distribution as a Gaussian convolved with an exponential function to represent the dust. The Skew-t function can be used to describe data with skewed and tail features. The Skew-t distribution is widely used for multivariate skew distributions in statistics, quantitative risk management, and insurance. We choose skew functions instead of Gaussian functions to model the posterior distributions of quasars. Details about the Skew-t function will be provided in \ref{subsec:skewt}.

In addition to the systematics of photo-$z$, quasar candidate selection is also a key issue. There are diverse methods used to select quasars. For example, the ultraviolet excess (UVX) method \citep{Sandage1965, Green1986} for $z<2.2$ quasars; X-ray sources \citep[e.g.,][]{Trump2009}; radio sources such as from the VLA FIRST survey \citep[e.g.,][]{Becker2000}; quasar variability \citep[e.g.,][]{NPD2011}; optical color box selection for the 2dF-SDSS LRG and QSO Survey \citep[2SLAQ,][]{Croom2009}, and for the SDSS target selection \citep[e.g.,][]{Richards2002}; more complex methods with optical (and infrared photometry), including non-parametric Bayesian classification and Kernel Density Estimator \citep[KDE,][]{Richards2004, Richards2009b}, XDQSO \citep{Bovy2011}, the neural network approach \citep{Yeche2010}, the Bayesian likelihood approach \citep{Kirkpatrick2011}; and selection combining different methods \citep[e.g.,][]{Ross2012}. When a survey goes fainter, the contamination of point-like galaxies becomes significant, with the contamination rate as a function of magnitude. Fitting a training sample with all point-like objects is not efficient with regard to quasar selection at different magnitudes. A training sample consisting of all point-like objects will include stars, quasars, and point-like galaxies, thus it is hard to fit their posterior distribution all together. To separate quasars from stars, we estimate the number counts and colors (relative fluxes) of stars from a Milky Way synthetic simulation with the Besan\c{c}on model. We also do galaxy template fitting to help distinguish galaxies from quasars.

The paper is organized as follows. In Section \ref{sec:data}, we introduce the spectroscopically identified quasar sample and photometric data used in this work. In Section \ref{sec:method}, we describe the photo-$z$ regression algorithm. We compare the photo-$z$ results obtained by different photo-$z$ methods using the same optical photometric data in Section \ref{sec:results}. We also present photo-$z$ results using SDSS, SDSS-WISE, PS1-WISE and DECaLS-WISE photometry in Section \ref{sec:results}. We present the classification method in Section \ref{sec:classification}, including the stellar simulation, the galaxy template fitting, and the Bayesian classification method. Quasar candidate selection using the DECaLS and WISE photometry is presented in Section \ref{sec:discussion}. We test the results in some deep fields and present the quasar number count statistics in the SDSS Stripe 82 (S82) region. We summarize the paper in Section \ref{sec:summary}. We will make the photo-$z$ and classification code publicly available\footnote{\url{https://github.com/qian-yang/Skewt-QSO}} with the current version archived in Zenodo \citep{SkewtQSOv1}. In the paper, all magnitudes are expressed in the AB system. The galactic extinction of extragalactic objects is corrected using the dust reddening map of \citet{Schlegel1998}. We discuss only type 1 quasars (or AGNs) in this work. We use a $\Lambda$CDM cosmology with $\Omega_{\Lambda}=0.728$, $\Omega_m=0.272$, $\Omega_b=0.0456$, and $H_0=70$ km s$^{-1}$ Mpc$^{-1}$ \citep{Komatsu2009}.

\section{The Data} \label{sec:data}
\subsection{Spectroscopically Identified Quasar Sample} \label{subsec:sample}
We use a sample of spectroscopically identified quasars consisting of quasars from the SDSS Data Release 7 Quasar catalog (DR7Q) \citep{Schneider2010} and the Data Release 12 Quasar catalog (DR12Q) \citep{Paris2017}. There are 105,783 quasars in the DR7Q, and 297,301 quasars in the DR12Q, including 25,275 quasars in both catalogs. BAL quasars are anomalously redder than most quasars and are excluded from our analysis. There are 29,580 quasars identified as BAL quasars in the DR12Q, and 6,214 quasars in the DR7Q identified as BAL quasars by \citet{Shen2011}. After removal of the BAL quasars, there are 346,464 quasars in our quasar sample (DR7\&12). Since, in comparison with the SDSS photometric bands, there are more high redshift quasars detected by the redder PS1 $y$ band and deeper DECaLS $z$ band, it is now possible to construct color models for high redshift quasars. We also include some quasars, which are not in the SDSS DR7 or DR12 catalog. A high redshift quasar catalog with 437 $z>4.5$ (called the BONUS high redshift sample) was constructed from the literature (Table 1 and Table 3 in \citet{Wang2015} and references therein; Table 7 in \citet{Banados2016} and references therein; \citealt{Jiang2016}; \citealt{Yang2017}; \citealt{Wang2017}).

\subsection{SDSS Photometry} \label{subsec:sdss}
We use the point spread function \citep[PSF;][]{Lupton1999} photometry in the five SDSS bands $ugriz$ \citep{Fukugita1996}. The magnitude limits
(95\% completeness for point sources) in the five bands are 22.0, 22.2, 22.2, 21.3, and 20.5 mag, respectively. We queried the photoObjAll table in the SDSS CASJOB, and got the SDSS photometry for 304,241 quasars with restrictions on mode and flags \citep{Stoughton2002, Bovy2011, Richards2015}. The Galactic extinction coefficients for E(B-V) used are $Au, Ag, Ar, Ai, Az = 5.155, 3.793, 2.751, 2.086, 1.479$. The $u$ band and $z$ band are converted to the AB system using $u_{AB} = u_{SDSS} - 0.04$ mag and $z_{AB} = z_{SDSS} + 0.02$ mag \citep{Fukugita1996}.

\subsection{PS1 Photometry} \label{subsec:ps1}
We use the PSF photometry in the PS1 survey. The 5$\sigma$ median limiting AB magnitudes in the five PS1 bands $grizy$ are 23.2, 23.0, 22.7, 22.1, and 21.1 mag, respectively. We queried the StackObjectThin table in the PS1 CASJOB with restrictions on primaryDetection and infoFlag, and got PS1 photometry for 344,318 quasars. Due to the difference between the absorbing column of the atmosphere at the two survey sites, the extinction coefficients for the SDSS and PS1 filters are different. The Galactic extinction coefficients for E(B-V) are $Ag, Ar, Ai, Az, Ay = 3.172, 2.271, 1.682, 1.322, 1.087$ \citep{Schlafly2011}.

\subsection{WISE Photometry} \label{subsec:wise}
WISE \citep{Wright2010} mapped the sky at 3.4, 4.6, 12, and 22 $\mu$m (W1, W2, W3, W4). The 5$\sigma$ limiting magnitudes of the ALLWISE catalog in W1, W2, W3 and W4 bands are 19.6, 19.3, 16.7 and 14.6 mag. We use only WISE W1 and W2 photometric data, because the other two bands are much shallower. Out of 346,464 quasars in the DR7\&12 spectroscopic quasar catalog, 261,614 (76\%) and 256,606 (74\%) quasars are detected within 2 arcseconds in the WISE ALLWISE W1 and W2 bands, respectively. The WISE magnitudes are converted from Vega magnitude to AB magnitude with $\Delta m = $ 2.699 and 3.339 for the W1 and W2 bands, respectively.

\subsection{DECaLS Photometry} \label{subsec:decals}
The DESI Legacy imaging survey (DELS; Dey et al. 2017, in preparation) will provide images for target selection, including the DECam Legacy Survey (DECaLS) in the $g$, $r$ and $z$ bands, the Beijing-Arizona Sky Survey \citep[BASS;][]{Zou2017} in the $g$ and $r$ bands, and the Mayall $z$-band Legacy Survey (MzLS). The 5$\sigma$ point-source magnitude limits in $g$, $r$, and $z$ will be roughly 24.7, 23.9, and 23.0 mag. With depths of $1.5-2.5$ mag fainter than in the SDSS, the DELS will be useful in searching for fainter quasars than the SDSS spectroscopic quasars, and also high redshift quasars \citep{Wang2017}. In this work, we use the three band ($grz$) photometry from the DECaLS DR3. There are 194,529 known quasars detected in the DECaLS DR3 catalogs, and 98,481 quasars observed in all three bands in DR3. There are 235 quasars in the BONUS high redshift sample detected in the DECaLS DR3, and 149 of them were observed in the $g$, $r$, and $z$ bands. The unWISE coadds the WISE imaging and has better resolution \citep{Lang2014}. The unWISE $5\sigma$ detection rates for our spectroscopic quasar sample are higher than those for WISE, 87.6\% and 77.1\% for the W1 and W2 bands, respectively. The unWISE photometry is available in the DECaLS catalogs. For objects with detections lower than 5$\sigma$, the unWISE data are still included in the DECaLS catalogs with corresponding larger photometric errors. We use the unWISE W1 and W2 band photometry, instead of ALLWISE, when using the DECaLS optical photometry.

\begin{figure}[htbp]
\hspace*{-0.3cm}
\epsscale{1.2}
\plotone{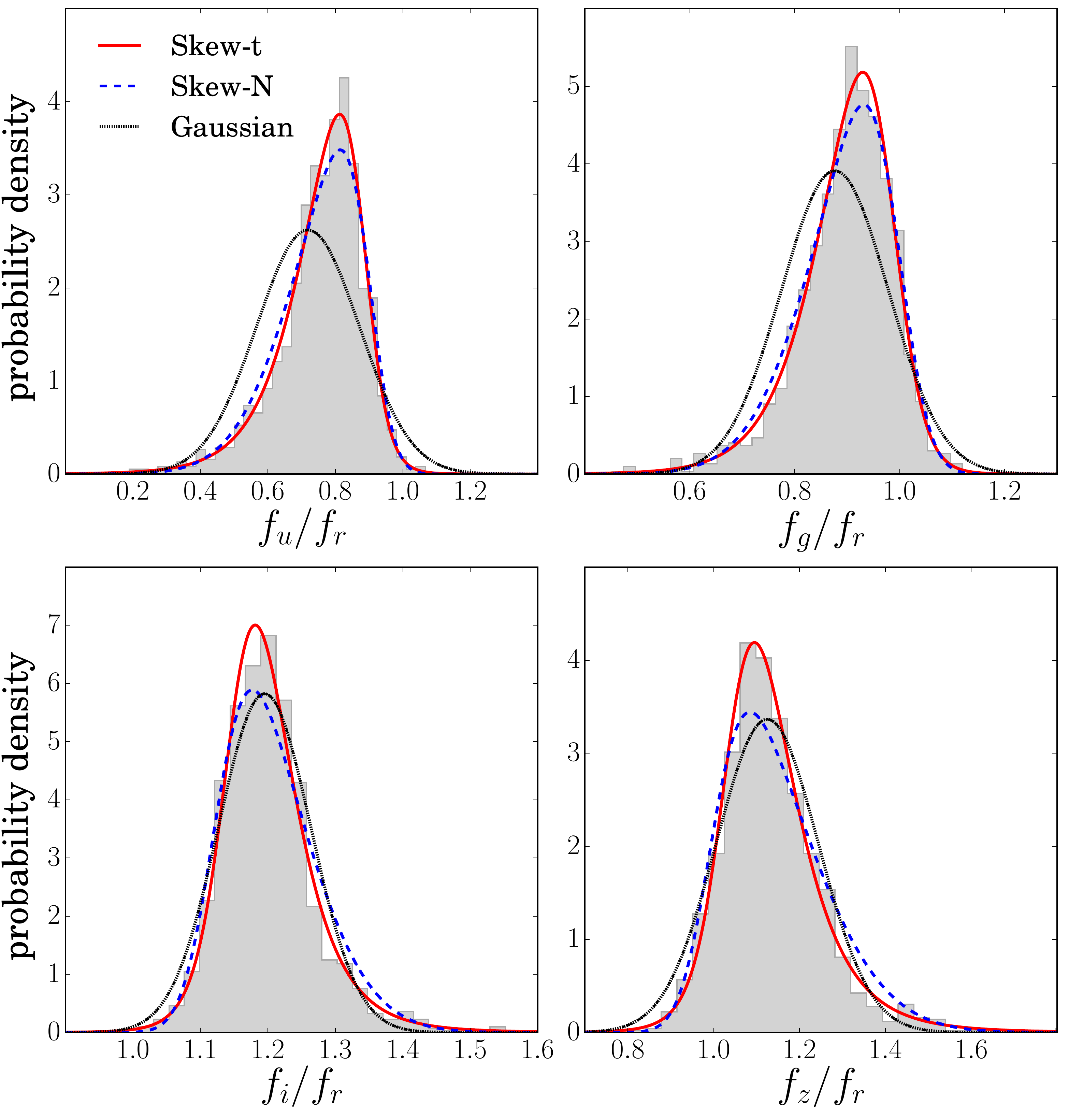}
\caption{\label{fig:skewt}Examples of one dimensional relative flux $f_u/f_r$, $f_g/f_r$, $f_i/f_r$ and $f_z/f_r$ distributions for quasars with $1.5<z_s<1.6$ and $18.5<r<19.0$. The relative flux distributions are skewed and show tail features even in a small redshift and magnitude bin. Obviously, the Skew-t model (red solid line) fits the relative flux distributions better than the Gaussian model (black dotted line) and the Skew-Normal model (blue dashed line).}
\end{figure}
\section{The photo-$z$ regression algorithm} \label{sec:method}
\subsection{Posterior probability with Multivariate Skew-t model} \label{subsec:skewt}
In this work, we model the posterior probability distribution for the relative fluxes of quasars using multivariate Skew-t distributions, with (1) ``skew" considering the asymmetric characteristic; (2) ``t" (student distribution) considering the incompleteness of the spectroscopically identified quasar sample. 

Multivariate skew-normal densities extend the multivariate normal model by allowing a shape parameter to account for skewness \citep{Azzalini1985, Azzalini1996}. The probability density function (PDF) of an n-dimensional multivariate skew-normal distribution is $SN_n(\bm{\mu}, \bm{\Sigma}, \bm{\lambda)}$, where $\bm{\mu}$ is the mean vector, $\bm{\Sigma}$ is the covariance matrix, and $\bm{\lambda}$ is the shape parameter vector. The distribution can be written as
\begin{equation} \label{eq:1}
 2\phi_n(\mathbf{x}|\bm{\mu}, \bm{\Sigma})\Phi(\bm{\lambda}^{\rm T}\bm{\Sigma}^{-1/2}(\mathbf{x}-\bm{\mu})),
\end{equation}
where $\phi_n(\mathbf{x}|\bm{\mu}, \bm{\Sigma})$ is the PDF of the n-variate normal distribution, ${\bm{\lambda}}^{\rm T}$ is the transform vector of $\bm{\lambda}$, and $\Phi(\bm{\lambda}^{\rm T}\bm{\Sigma}^{-1/2}(\mathbf{x}-\bm{\mu}))$ is the cumulative distribution function (CDF) of the standard normal distribution. When $\bm{\lambda} = 0$, the skew normal distribution becomes the normal distribution $ N_n(\bm{\mu}, \bm{\Sigma}$).

The Student-t distribution is used to estimate the mean of a normally distributed population when a sample size is small and its standard deviation is unknown. Adding a parameter for the number of degrees of freedom $\nu$, the PDF of the multivariate student-t distribution can be expressed as \citep[e.g.,][]{Johnson1994, Lachos2014},
\begin{equation} \label{eq:2}
  \frac{\Gamma(\frac{n+\nu}{2})}{\Gamma(\frac{\nu}{2})(\nu\pi)^{n/2}}|\bm{\Sigma}|^{-1/2}(1+\frac{d}{\nu})^{-(\frac{n+\nu}{2})},
\end{equation}
where $\Gamma$ is the gamma function, and d is the Mahalanobis distance $d = (\mathbf{x} - \bm{\mu})^{\rm T}\bm{\Sigma}^{-1}(\mathbf{x} - \bm{\mu})$.
When $\nu = \infty$, the student-t distribution becomes the the normal distribution.

With a shape parameter vector $\bm{\lambda}$ and a degree of freedom parameter $\nu$, the PDF of the multivariate Skew-t distribution $ST_n(\bm{\mu}, \bm{\Sigma}, \bm{\lambda}, \nu)$ can be described as \citep[e.g.,][]{Johnson1994, Lachos2014},
\begin{equation} \label{eq:3}
  2t_n(\mathbf{x}|\bm{\mu}, \bm{\Sigma}, \nu)T_n(\mathbf{x}|\bm{\mu}, \bm{\Sigma}, \nu),
\end{equation}
where $t_n(\mathbf{x}|\bm{\mu}, \bm{\Sigma}, \nu)$ and $T_n(\mathbf{x}|\bm{\mu}, \bm{\Sigma}, \nu)$ are the PDF and CDF of the student-t distribution. When $\nu = \infty$ the Skew-t distribution becomes the Skew-normal distribution. Figure \ref{fig:skewt} shows the distributions of quasar relative fluxes $f_u/f_r$, $f_g/f_r$, $f_i/f_r$, $f_z/f_r$ in a narrow bin with $1.5<z_s<1.6$ and $18.5<r<19.0$. These relative flux distributions are obviously skewed and show tails even when excluding broad absorption line (BAL) quasars. We present the Anderson-Darling goodness of fit tests \citep{Marsaglia2004} using the R $ADGofTest$ package\footnote{https://cran.r-project.org/web/packages/ADGofTest} for distributions shown in Figure \ref{fig:skewt}. The probability values fit with the Gaussian function, the Skew-normal function, and the Skew-t function for $f_u/f_r$ distribution are $4.46\mathrm{e}{-7}$, 0.0168, and 0.768; for $f_g/f_r$ distribution are $4.46\mathrm{e}{-7}$, 0.111, and 0.986; for $f_i/f_r$ distribution are $4.87\mathrm{e}{-6}$, $5.48\mathrm{e}{-4}$, and 0.539; and for $f_z/f_r$ distribution are $4.47\mathrm{e}{-7}$, $7.72\mathrm{e}{-5}$, and 0.913. This indicates clearly that the advantage of Skew-t model is statistically significant. The Skew-t functions fit the quasar relatve fluxes better than the Gaussian and the Skew-Normal functions do.

The quasar sample is divided into a grid of redshifts and magnitudes with bin sizes of $\Delta z = 0.05$ and $\Delta m = 0.1$. A redshift bin 0.05 is acceptable for photo-$z$ regression. A smaller redshift bin will lead to poor statistics in any single bin. We use the R $sn$ package\footnote{https://cran.r-project.org/web/packages/sn/index.html}$^{,}$\footnote{http://azzalini.stat.unipd.it/SN/} to do a maximum penalized likelihood estimation to model the multivariate relative flux distribution in each redshift and magnitude bin, and get the $\bm{\mu}(z, m)$, $\bm{\Sigma}(z, m)$, $\bm{\lambda}(z, m)$ and $\nu(z, m)$ parameters. The $g$-band magnitude of the SDSS reaches a fainter depth than other bands, and the second faint band is the $r$ band \citep{McGreer2013}. Due to the Lyman-$\alpha$ emission shifting out of the $g$ band and the Lyman forest absorptions, g-band magnitudes of $z>4.6$ quasars become faint. Quasars are therefore divided into magnitude bins based on the $r$-band magnitude, and we use the relative fluxes between other band fluxes and the $r$-band flux, for example $f_u/f_r$, $f_g/f_r$, $f_i/f_r$ and $f_z/f_r$ when using the SDSS five-band photometry. Each relative flux is a dimension in the Skew-t multi-dimensional model. The covariance between relative fluxes is accounted for by the covariance matrix $\bm{\Sigma}$.

To calculate the PDF, we weigh relative fluxes using photometric uncertainties as follows. For example, for the four relative fluxes of the SDSS photometry $f_u/f_r$, $f_g/f_r$, $f_i/f_r$ and $f_z/f_r$, with flux uncertainties $e_{u}$, $e_{g}$, $e_{r}$, $e_{i}$ and $e_{z}$ in the five SDSS bands, the relative flux covariance matrix can be derived from the error propagation equations as,
\begin{equation} \label{eq:4}
\bm{\Sigma}_0 =
\left(
\begin{array}{ccccc}
\frac{e_u^2 f_r^2 + e_r^2 f_u^2}{f_r^4} & \frac{f_u f_g e_r^2}{f_r^4} & \frac{f_u f_i e_r^2}{f_r^4} & \frac{f_u f_z e_r^2}{f_r^4} \\
\frac{f_u f_g e_r^2}{f_r^4} & \frac{e_g^2 f_r^2 + e_r^2 f_g^2}{f_r^4} & \frac{f_g f_i e_r^2}{f_r^4} & \frac{f_g f_z e_r^2}{f_r^4} \\
\frac{f_u f_i e_r^2}{f_r^4} & \frac{f_g f_i e_r^2}{f_r^4} & \frac{e_i^2 f_r^2 + e_r^2 f_i^2}{f_r^4} & \frac{f_i f_z e_r^2}{f_r^4} \\
\frac{f_u f_z e_r^2}{f_r^4} & \frac{f_g f_z e_r^2}{f_r^4} & \frac{f_i f_z e_r^2}{f_r^4} & \frac{e_z^2 f_r^2 + e_r^2 f_z^2}{f_r^4}\\
\end{array}
\right)
.
\end{equation}
\vspace{0cm}

When combining optical photometry and mid-infrared photometry that are taken separately for years, the quasar variability introduces extra uncertainties into the relative fluxes, such as $f_{W1}/f_r$ and $f_{W2}/f_r$. To reduce the uncertainties from quasar variability, we use $f_{W1}/f_r$ and $f_{W2}/f_{W1}$ for quasar photo-$z$ estimation. In the case of using the DECaLS $g$, $r$, $z$ and WISE W1, W2 photometry, the relative fluxes used are $f_g/f_r$, $f_z/f_r$, $f_{W1}/f_r$ and $f_{W2}/f_{W1}$, and the covariance matrix $\bm{\Sigma}_0$ is written as
\begin{equation} \label{eq:5}
\left(
\begin{array}{ccccc}
\frac{e_g^2 f_r^2 + e_r^2 f_g^2}{f_r^4} & \frac{f_g f_z e_r^2}{f_r^4} & \frac{f_g f_{W1} e_r^2}{f_r^4} & 0\\
\frac{f_g f_z e_r^2}{f_r^4} & \frac{e_z^2 f_r^2 + e_r^2 f_z^2}{f_r^4} & \frac{f_z f_{W1} e_r^2}{f_r^4} & 0\\
\frac{f_g f_{W1} e_r^2}{f_r^4} & \frac{f_z f_{W1} e_r^2}{f_r^4} & \frac{e_{W1}^2 f_r^2 + e_r^2 f_{W1}^2}{f_r^4} & -\frac{e_{W1}^2 f_{W2}}{f_{W1}^2}\\
0 & 0 & -\frac{e_{W1}^2 f_{W2}}{f_{W1}^2} & \frac{e_{W2}^2 f_{W1}^2 + e_{W1}^2 f_{W2}^2}{f_{W1}^4} \\
\end{array}
\right).
\end{equation}

Then the covariance matrix is $\bm{\Sigma}^*(z, m) = \bm{\Sigma}(z, m) + \bm{\Sigma}_0$. The posterior probability is expressed as
\begin{equation} \label{eq:6}
P_{\rm QSO}(\bm{f}|z, m) = ST_n(\bm{\mu}(z, m), \bm{\Sigma^*}(z, m), \bm{\lambda}(z, m), \nu(z, m)),
\end{equation}
where $\bm{f}$ represents the relative fluxes.

PS1, DECaLS, and WISE photometry are based on multiple epochs of imaging data, and the effect of variability is mitigated. It happens
that PS1, DECaLS, and WISE are all averages over a roughly similar timeframe (although DECaLS is mostly a couple of years after PS1),
whereas SDSS is about a decade earlier than the others. Therefore, combinations of SDSS and the other surveys will be the most impacted
by long-term variability.

\begin{figure}[htbp]
\hspace*{-0.5cm}
\epsscale{1.3}
\plotone{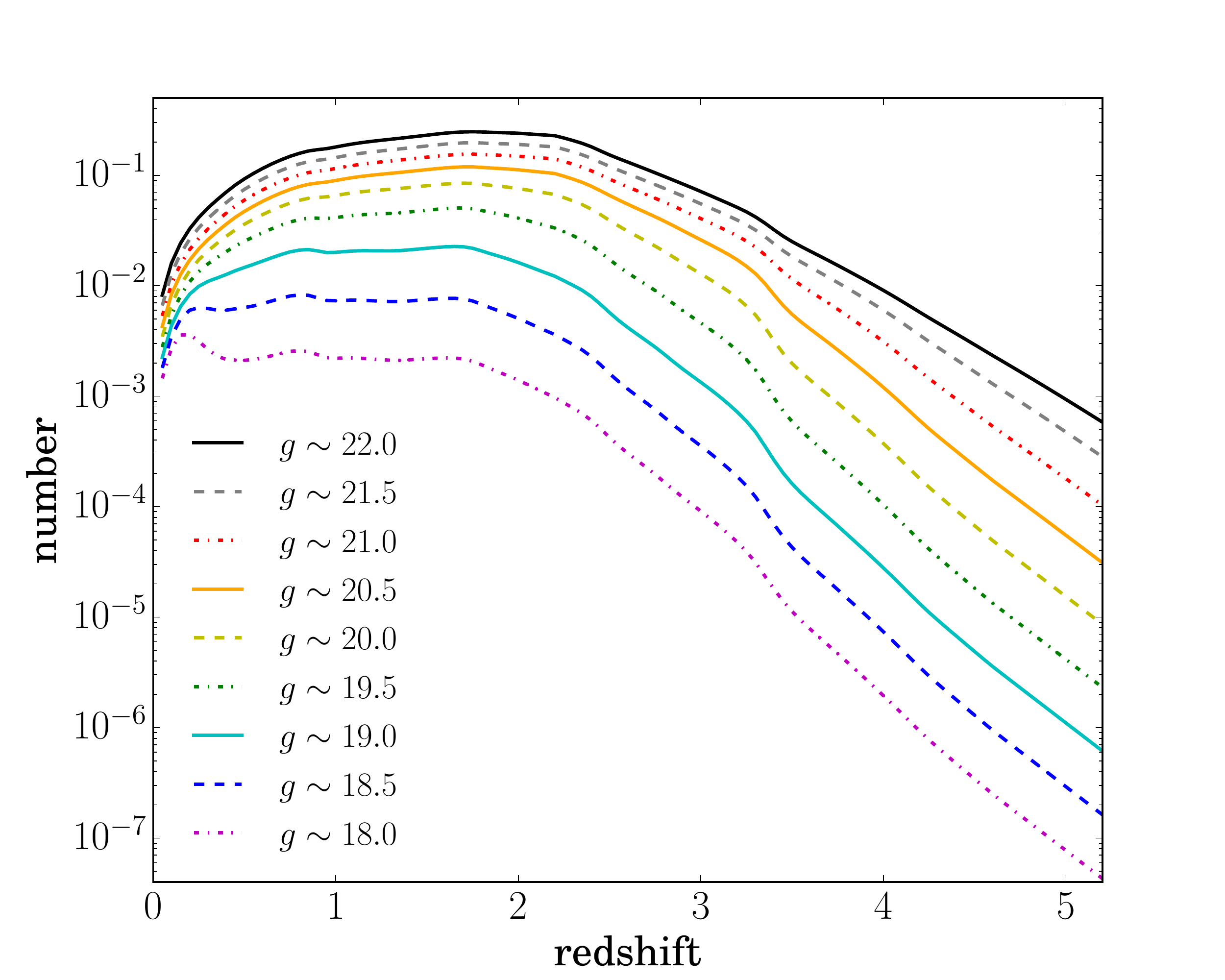}
\vspace{-0.5cm}
\caption{\label{fig:qlf}The quasar number prior $N_{\rm QSO}(z, m)$ per deg$^2$ as a function of redshift and magnitude with $\Delta z = 0.05$ and $\Delta g = 0.1$, derived from the QLF in \citet{NPD2016}. The curves, from top to bottom, are for $g = $ 22.0, 21.5, 21.0, 19.5, 19.0, 18.5, 18.0 mag, respectively.}
\end{figure}

\subsection{Prior Probability from the QLF} \label{subsec:qlf}
The number density of quasars depends on the redshift and luminosity \citep[e.g.,][]{Ross2013, NPD2016}. The QLF characterizes quasars through the evolution of their number density with luminosity and redshift. \citet{NPD2016} present the QLF using quasars from the extended Baryon Oscillation Spectroscopic Survey of the Sloan Digital Sky Survey (SDSS-IV/eBOSS). Their quasar sample is 80\% complete to $g = 20$ mag and 50\% complete to $g = 22.5$ mag, and the QLF has been corrected for incompleteness. We derive the quasar number prior $N_{\rm QSO}(z, m)$ per deg$^2$ as a function of redshift and magnitude with $\Delta z = 0.05$ and $\Delta g = 0.1$ from the QLF in \citet{NPD2016} derived in the SDSS $g$ band, and with the k-corrections as a function of both redshift and luminosity \citep{McGreer2013}. Figure \ref{fig:qlf} shows the number distribution as a function of redshift for quasars with $g = $ 22.0, 21.5, 21.0, 19.5, 19.0, 18.5, 18.0 mag from top to bottom.

Thus the PDF is obtained with the posterior probability and the number prior as
\begin{equation} \label{eq:7}
  P_{\rm QSO}(z) = P_{\rm QSO}(\bm{f}|z, m) N_{\rm QSO}(z, m).
\end{equation}
Using the PDF as a function of redshift, $P_{\rm QSO}(z)$, the photo-$z$ can be estimated by a maximum probability method or a peak recognition with maximum integrated probability. We identify peaks in a PDF curve using the $findpeaks$ function in R $pracma$ package\footnote{https://cran.r-project.org/web/packages/pracma/index.html}, and calculate the photo-$z$ as the peak with the largest integrated PDF within a redshift range $(z_1, z_2)$. A parameter $P_{\rm prob}$ describes the probability that the redshift locates within $(z_1, z_2)$ is
\begin{equation} \label{eq:8}
P_{\rm prob} = \frac{\int_{z_1}^{z_2} P_{\rm QSO}(z) dz}{\int P_{\rm QSO}(z) dz}
.
\end{equation}

The logarithmic likelihood ($L$) of an object to be a quasar over the whole redshift range is written as
\begin{equation} \label{eq:9}
L_{\rm QSO} = {\rm log}(P_{\rm QSO}) = {\rm log}\int P_{\rm QSO}(z) dz .
\end{equation}

To assess the impact of the prior distribution on the photo-$z$ regression results, we also present the results with photo-$z$ derived only from the posterior distribution. The PDF from the posterior probability is
\begin{equation} \label{eq:10}
  P_{\rm QSO}'(z) = P_{\rm QSO}(\bm{f}|z, m).
\end{equation}
The logarithmic likelihood of an object to be a quasar from the posterior distribution over the whole redshift range is written as
\begin{equation} \label{eq:11}
L'_{\rm QSO} = {\rm log}(P_{\rm QSO}') = {\rm log}\int P_{\rm QSO}'(z) dz .
\end{equation}
The influence of prior distribution on the photo-$z$ regression and quasar candidate selection is discussed in Sections \ref{subsec:photoz_infrared} and \ref{subsec:selection}.

\begin{figure}[htbp]
\hspace*{-0.5cm}
\epsscale{1.3}
\plotone{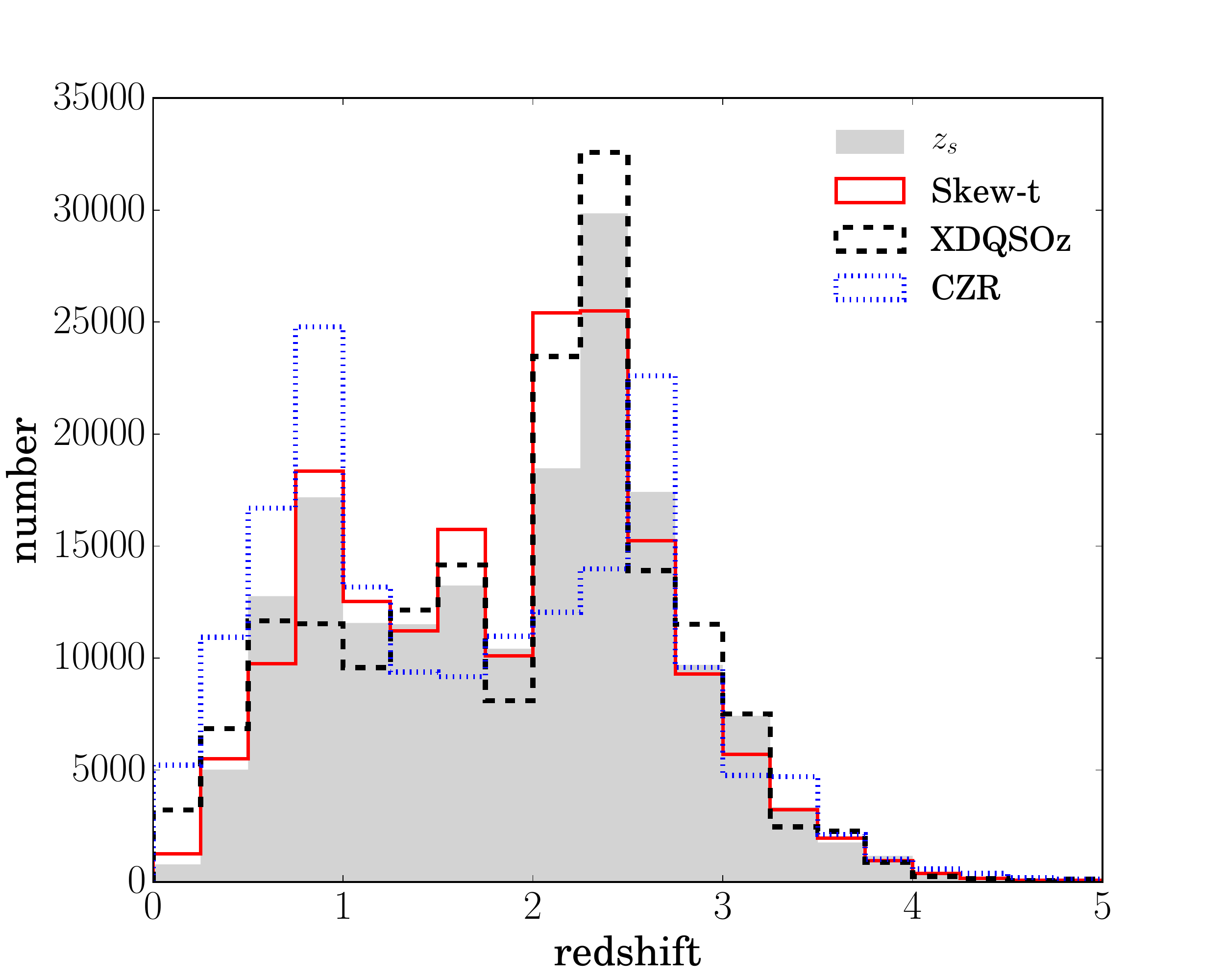}
\vspace{-0.5cm}
\caption{\label{fig:compare}The photo-$z$ distributions from different photo-$z$ methods, including Skew-t (red solid), XDQSOz (blue dashed) and CZR (blue dotted), compared with the spectroscopic redshift $z_s$ distribution (gray shade), using the SDSS five-band photometry for the same quasar test sample. The photo-$z$ distribution from the Skew-t model is more similar to the $z_s$ distribution, while the CZR method identifies more $z\sim 0.8$ quasars and the XDQSOz method identifies more $z \sim 2.2$ quasars.}
\end{figure}

\begin{figure*}
  \centering
  \hspace{0cm}
    \subfigure{
    \includegraphics[width=3.6in]{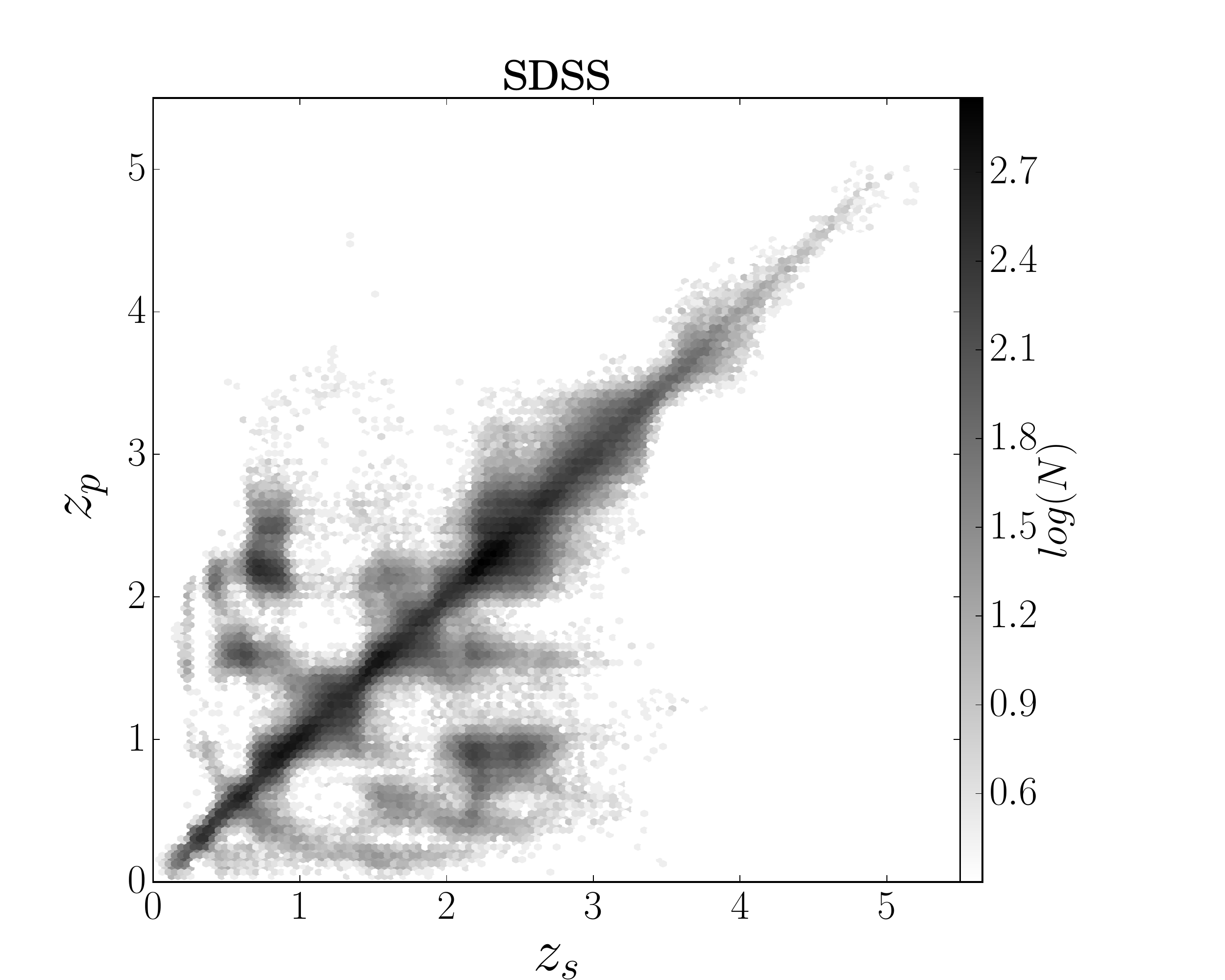}}
  \hspace{-1.4cm}
  \subfigure{
    \includegraphics[width=3.6in]{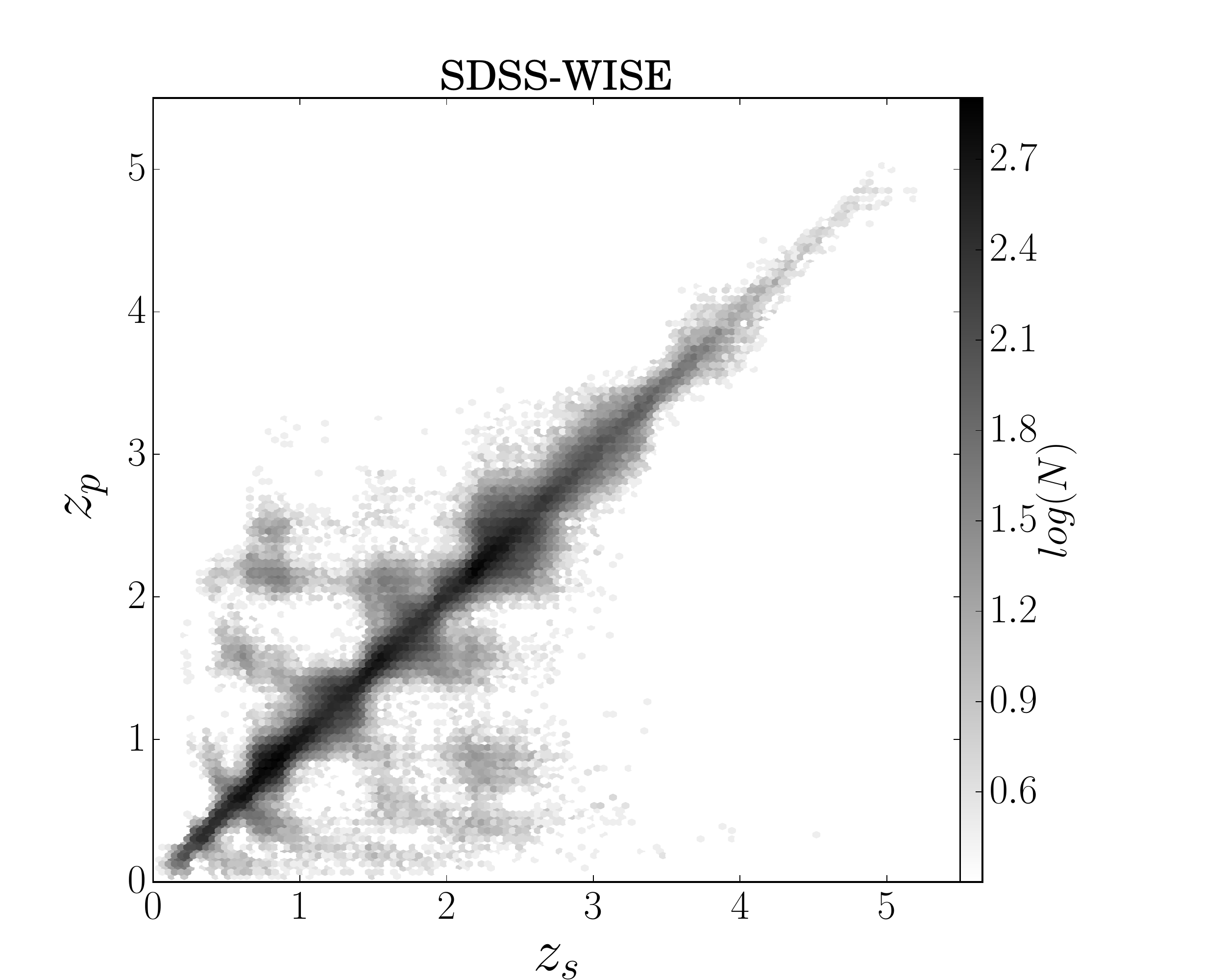}}\\
  \vspace{-0.3cm}
  \hspace{0cm}
  \subfigure{
    \includegraphics[width=3.6in]{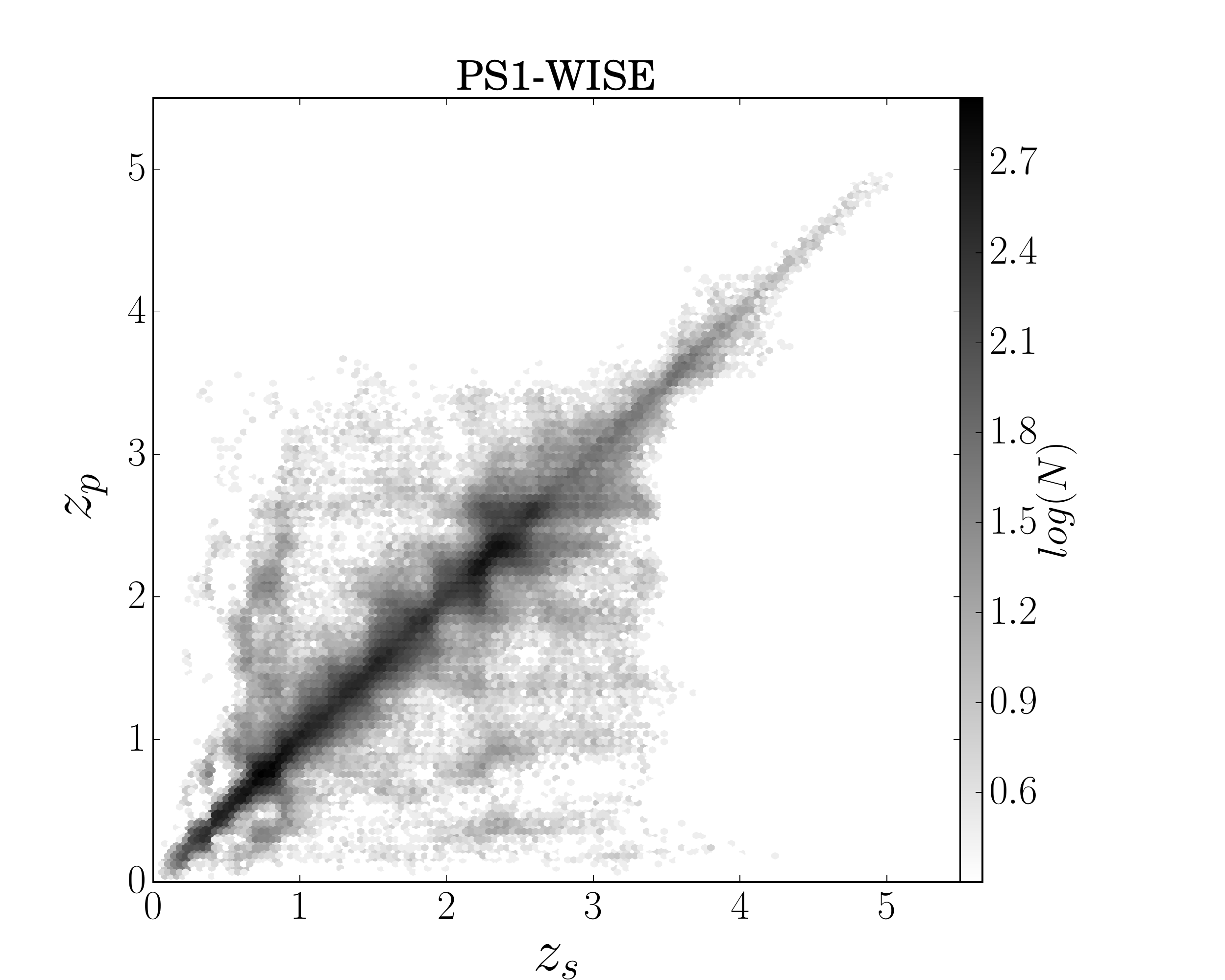}}
  \hspace{-1.4cm}
  \subfigure{
    \includegraphics[width=3.6in]{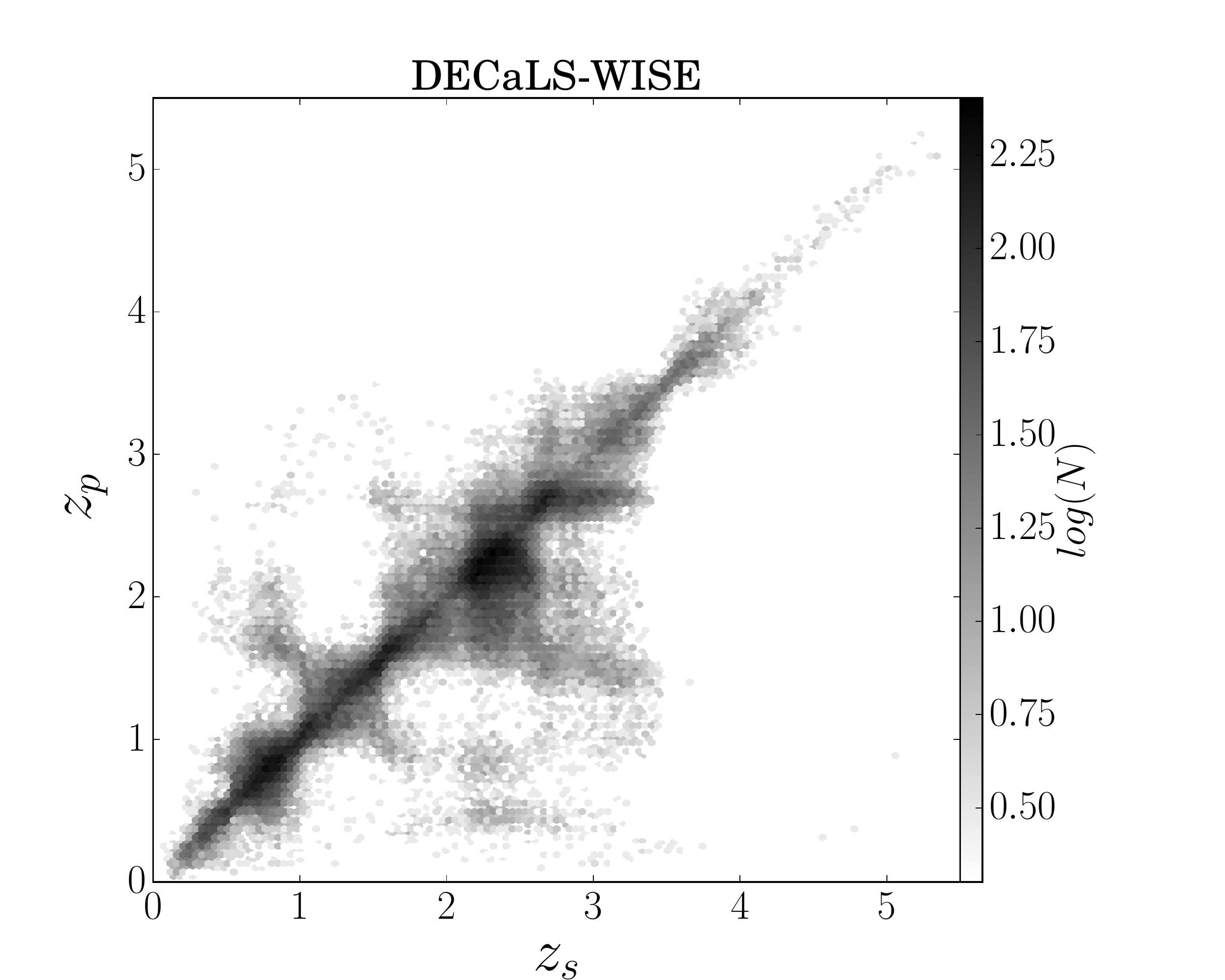}}\\
  \caption{\label{fig:photoz}Photo-$z$ ($z_p$) compared with spectroscopic redshifts ($z_s$) for SDSS, SDSS-WISE, PS1-WISE and DECaLS-WISE photometry, respectively. The degeneracy between $z\sim 0.8$ and $z\sim2.2$ is obvious when using only the SDSS photometry, and is alleviated by combining optical data with mid-infrared photometry.}
  \vspace{0.5cm}
\end{figure*}

\begin{figure}[htbp]
\hspace*{-1cm}
\epsscale{1.4}
\plotone{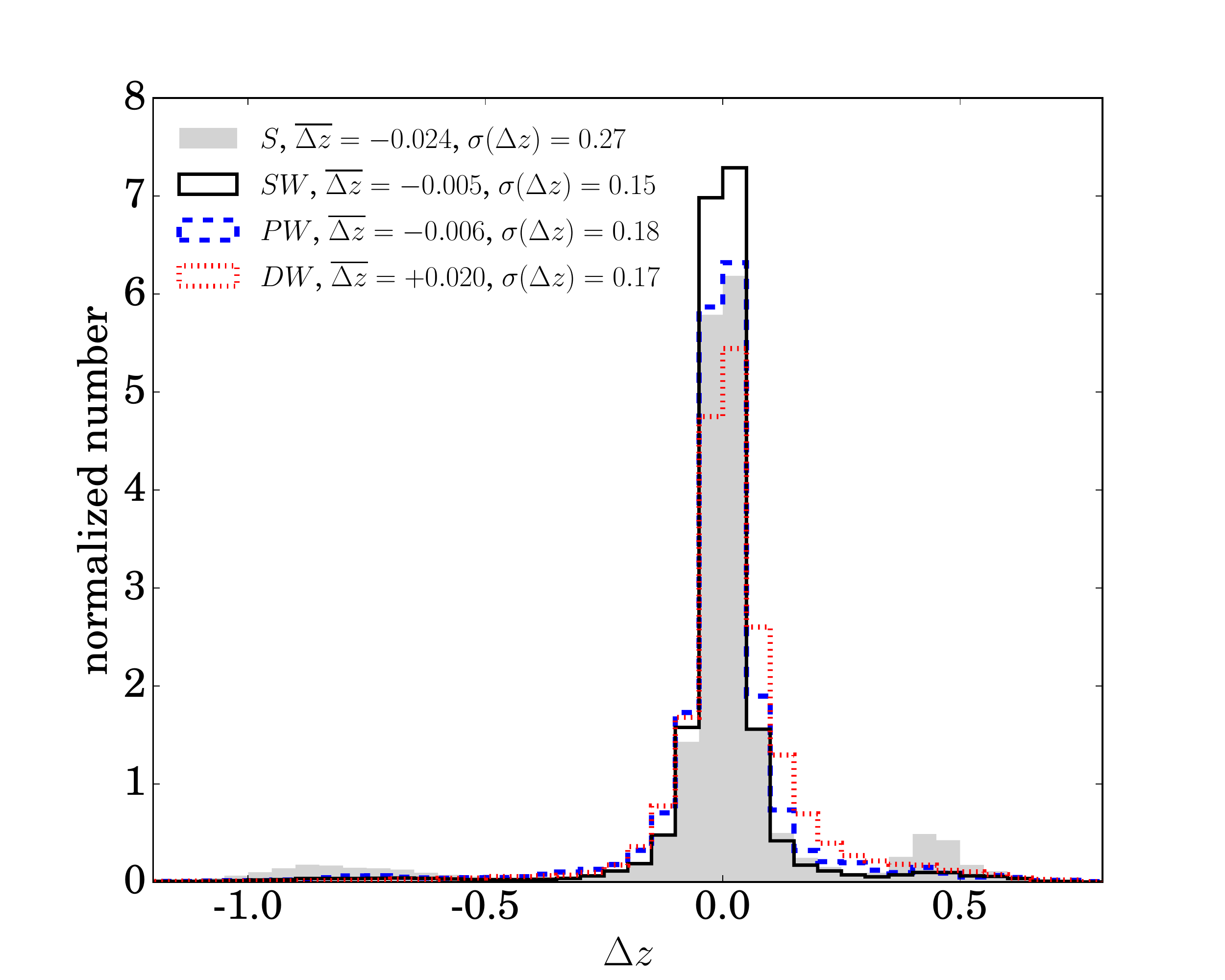}
\vspace{-0.5cm}
\caption{\label{fig:offset}The normalized distributions of $\Delta z$ for the different sources of photometry listed in Table \ref{tab:photoz_infrared}. The histograms are SDSS (S, gray shade), SDSS-WISE (SW, blue solid), PS1-WISE (PW, black dashed), and DECaLS-WISE (SW, red dotted), respectively.}
\end{figure}

\begin{figure*}
  \centering
  \hspace{-0.7cm}
    \subfigure{
    \includegraphics[width=3.7in]{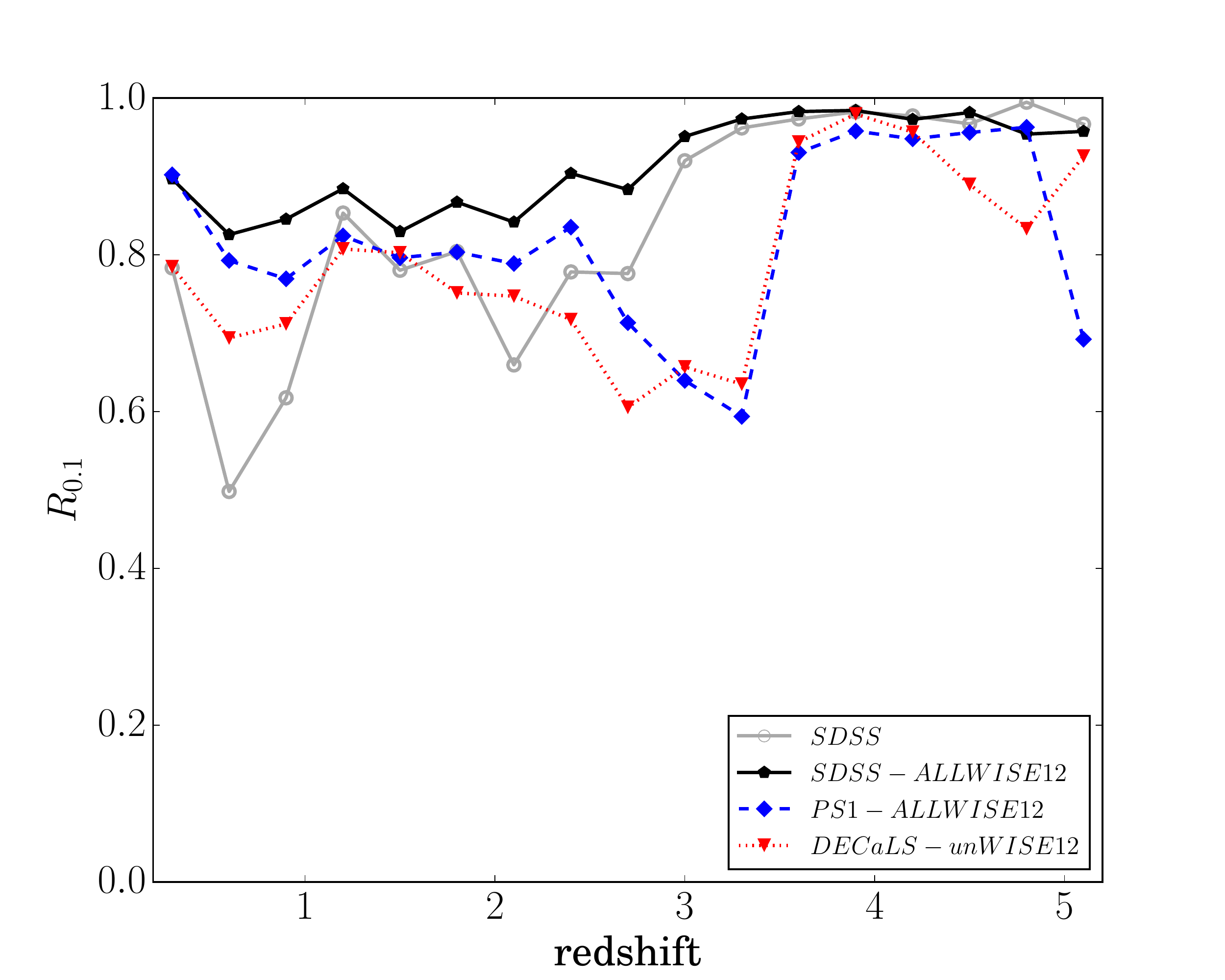}}
  \hspace{-1.2cm}
  \subfigure{
    \includegraphics[width=3.7in]{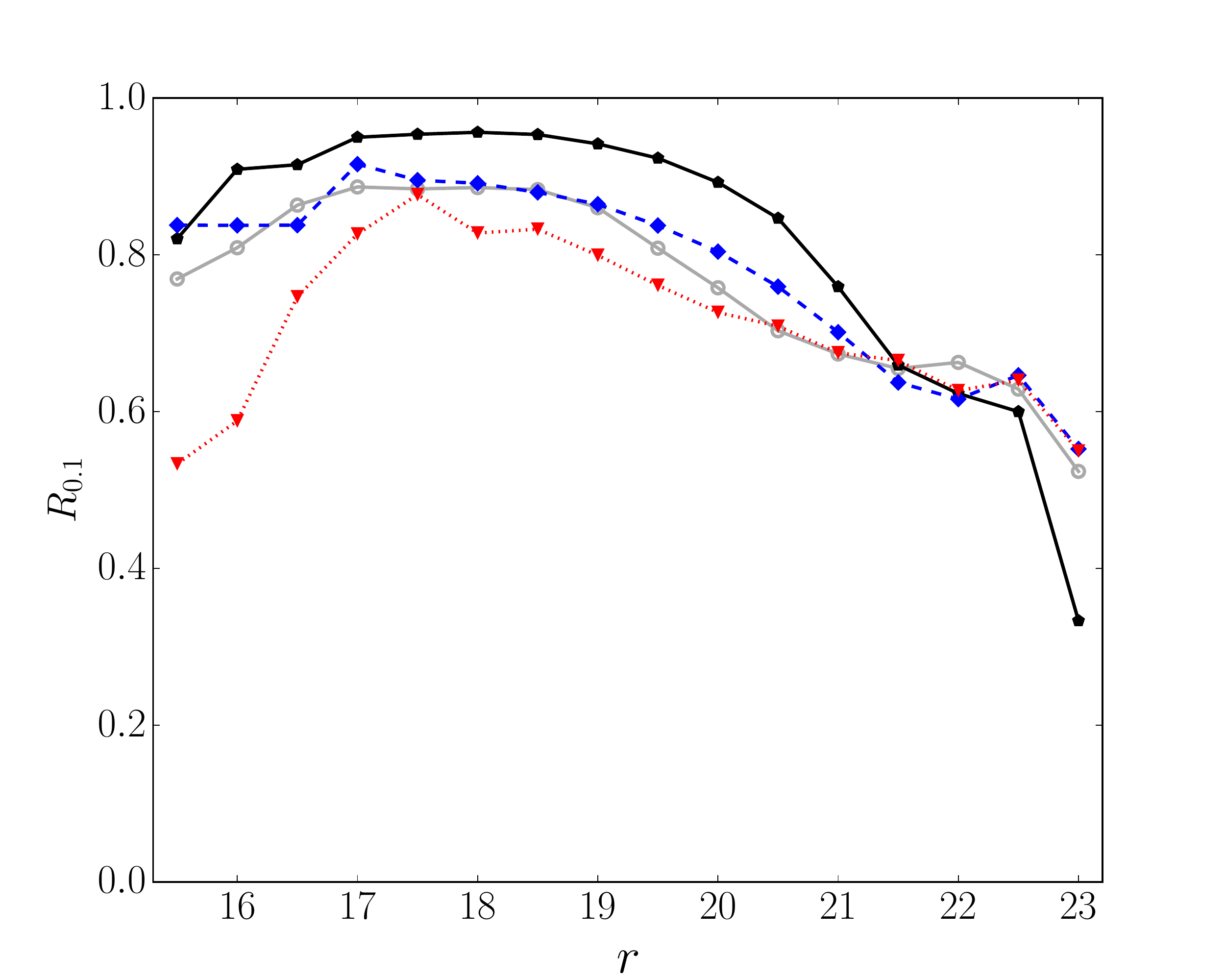}}
  \caption{\label{fig:raito}Photo-$z$ accuracy $R_{0.1}$ as a function of redshift (left panel) and magnitude (right panel) for different combinations of photometry, including SDSS (gray open dot-line), SDSS-WISE (black pentagon-line), PS1-WISE (black diamond-dashed line) and DECaLS-WISE (red triangle-dotted line). The photo-$z$ accuracies of PS1-WISE and DECaLS-WISE are lower than that of SDSS-WISE, mainly due to lack of $u$ band photometry in PS1 and DECaLS.} 
\end{figure*}

\begin{table}
\small
\begin{center}
\tablewidth{1pt}
\caption{Photo-$z$ results for different methods with the same test sample of SDSS photometry} \label{tab:photoz}
\begin{tabular}{lcrcccccc}
\tableline\tableline
${\rm Method}$ & $\sigma(\Delta z)$ & $\overline{\Delta z}$ & $R_{0.1}$ & $R_{0.2}$ & ${\rm Time}$\\
\tableline
Skew-t & 0.27 & -0.02  & 74.2\% & 81.5\% & 1 \\
XDQSOz & 0.31 & -0.04 & 72.8\% & 79.3\% & 17 \\
KDE & 0.35 & -0.06 & 70.6\% & 77.9\% & 0.002 \\
CZR & 0.29 & 0.05 & 68.0\% & 73.9\% & 0.005  \\
\tableline
\end{tabular}
\tablecomments{$R_{0.1}$ ($R_{0.2}$) is the fraction of quasars with $|\Delta z|$ smaller than 0.1 (0.2). Time is calculated by using the same machine, and the time used by the Skew-t method to obtain the photo-$z$ results for the test sample is normalized to 1. A test calculation of 100,000 objects with SDSS five bands data using Skew-t method took 23 minutes (by one processor computer with 3.1 GHz Intel Core i7 CPU).}
\end{center}
\end{table}

\begin{table*}
\small
\begin{center}
\tablewidth{1pt}
\caption{Photo-$z$ results for different sources of photometry} \label{tab:photoz_infrared}
\begin{tabular}{ccccrccc}
\tableline\tableline
${\rm Photometry}$ & $N_{\rm bands}$ & ${\rm PDF}$ & $\sigma(\Delta z)$ & $\overline{\Delta z}$ & $R_{0.1}$ & $R_{0.2}$ & ${\rm Numbe}r$ \\
\tableline
SDSS & 5 & posterior \& prior  & 0.27 & -0.024  & 74.9\% & 82.0\% &304,241 \\
- & -  & posterior & 0.29 & -0.025  & 73.5\% & 81.1\% & - \\
\tableline
SDSS-WISE & 7 & posterior \& prior & 0.15 & -0.005 & 87.0\% & 93.3\% & 229,653 \\
- & - & posterior & 0.16 & -0.007 & 85.8\% & 92.5\% & - \\
\tableline
PS1-WISE & 7 & posterior \& prior & 0.18 & -0.006 & 79.1\% & 89.5\% & 254,349 \\
- & - & posterior & 0.22 & -0.03 & 77.0\% & 87.6\% & - \\
\tableline
DECaLS-WISE & 5 & posterior \& prior & 0.17 & 0.020 & 72.4\% & 88.0\% & 98,450 \\
- & - & posterior & 0.23 & -0.002 & 72.3\% & 87.5\% & - \\
\tableline
\end{tabular}
\tablecomments{The photo-$z$ results are calculated with the PDF derived from the posterior and prior distributions in Equation (\ref{eq:7}) or from the posterior distribution in Equation (\ref{eq:10}).}
\end{center}
\end{table*}

\section{photo-$z$ Results} \label{sec:results}
\subsection{Comparing Photo-$z$ Results using the SDSS Photometric Data} \label{subsec:photoz_sdss}
We compare the performance of our photo-$z$ regression algorithm with other methods by testing with the same sample of photometric data. We randomly divide the quasar sample with the SDSS photometric data into two subsamples, one as the training sample, and the other one as the test sample. We also try a KDE method mapping the two dimensional color-color distributions of $u-g$ versus $g-r$, $g-r$ versus $r-i$, and $r-i$ versus $i-z$ in redshift bins. The KDE photo-$z$ code is based on the KDE method in \citet{Silverman1986}. The CZR photo-$z$ is calculated based on the CZR method in \citet{Weinstein2004}. The XDQSOz photo-$z$ is calculated with the XDQSOz code \citep{Bovy2012}. The photo-$z$ results of the Skew-t, XDQSOz, KDE and CZR methods are listed in Table \ref{tab:photoz}.

The Skew-t photo-$z$ algorithm performs well compared with other photo-$z$ methods. The difference between the photo-$z$ ($z_p$) and the spectroscopic redshift ($z_s$) is expressed as $\Delta z = (z_s-z_p)/(1+z_s)$. $R_{0.1}$ ($R_{0.2}$) is the fraction of quasars with $|\Delta z|$ smaller than 0.1 (0.2). The standard deviation of $\Delta z$, $\sigma(\Delta z)$, from the Skew-t photo-$z$ is 0.27, slightly better than 0.31 and 0.29 from the XDQSOz and CZR methods. The KDE method is memory consuming when the number of dimensions is high. The KDE method also strongly depends on its training sample. When the test sample is the same as the training sample, the $R_{0.1}$ is as high as 85\% if 4 SDSS colors are used. It decreases to 70\% when the test sample is different from the training sample. So, the KDE photo-$z$ method is easily over-trained. For the Skew-t method, when the test sample is the same as the training sample, the accuracy $R_{0.1}$ changes by less than 1\% (74.2\% in Table \ref{tab:photoz} and 74.9\% in Table \ref{tab:photoz_infrared}).

\begin{figure}[htbp]
\hspace*{-1cm}
\epsscale{1.4}
\plotone{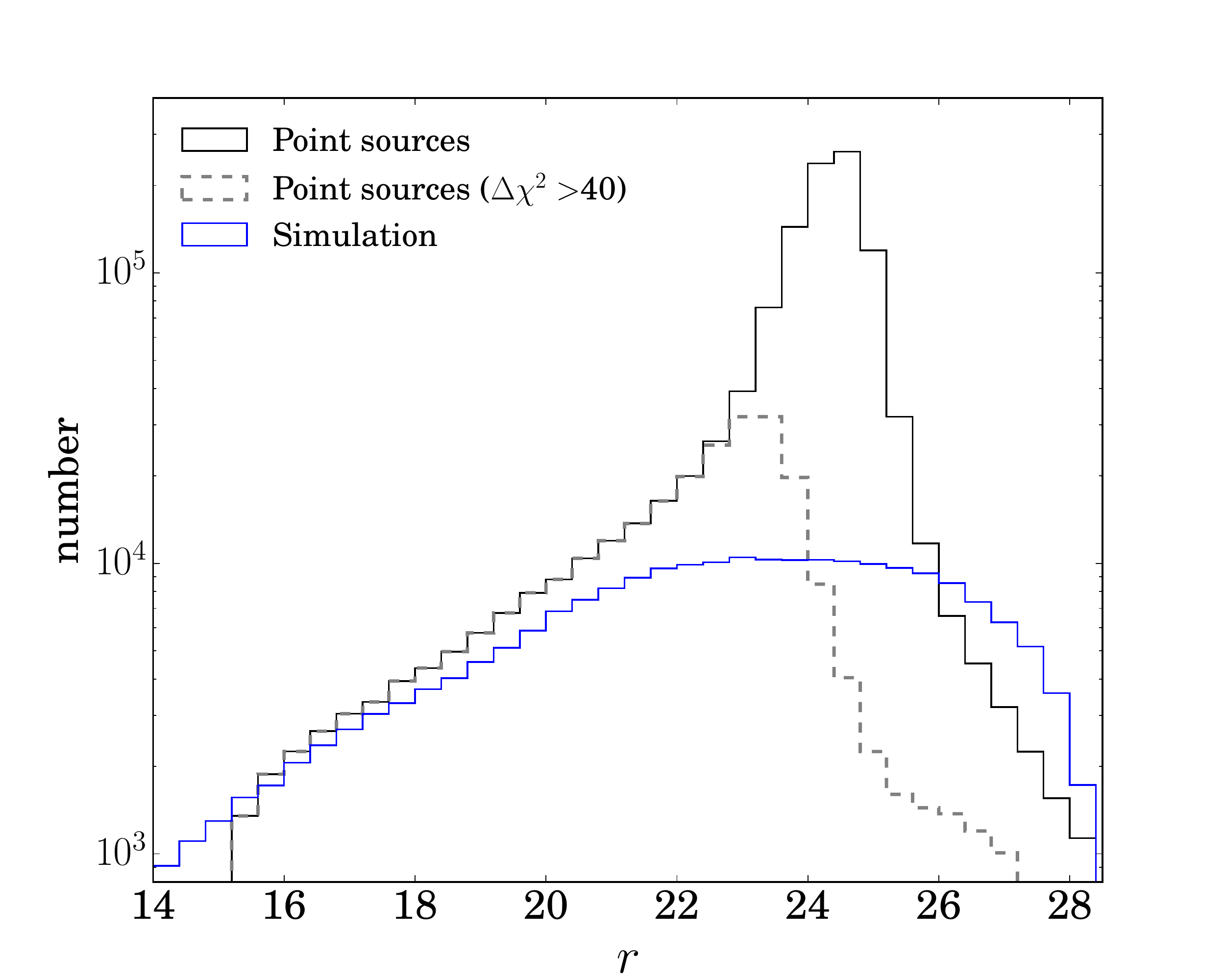}
\vspace{-0.5cm}
\caption{\label{fig:hist}The number distribution as a function of the $r$-band magnitude for the DECaLS ``PSF" type objects (black), ``PSF" type objects with $\Delta \chi^2>40$ (gray dashed), and stellar simulation (blue) within a 20 deg$^2$ test region in S82 ($340^{\circ}<{\rm R.A.}<350^{\circ}$, $-1^{\circ}<{\rm Decl.}<1^{\circ}$). The contaminations of point-like galaxies become prominent at the faint end.}
\end{figure}

\begin{figure*}
  \centering
  \hspace{-0.7cm}
    \subfigure{ 
    \includegraphics[width=3.5in]{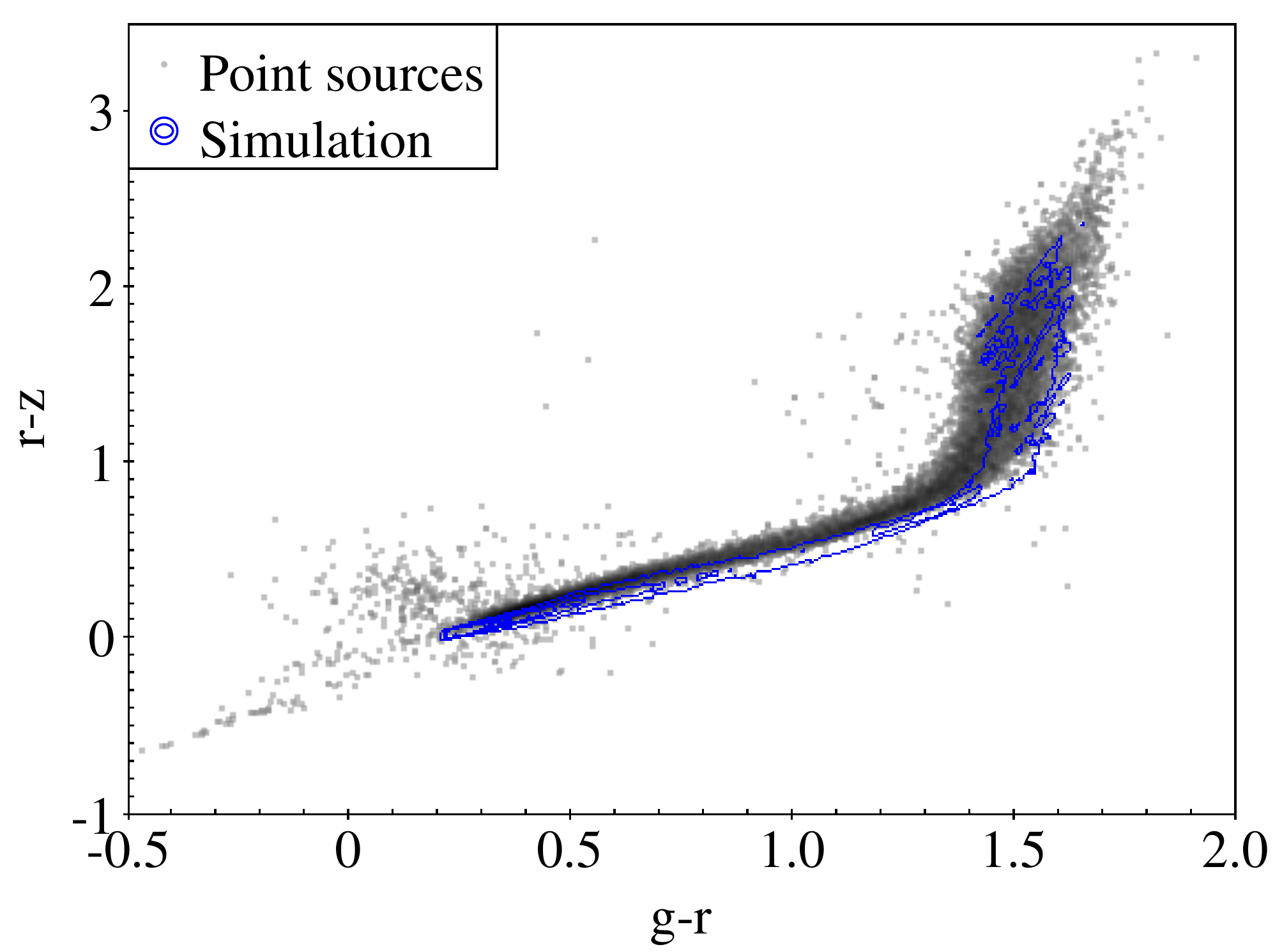}}
  \hspace{-0.2cm}
  \subfigure{ 
    \includegraphics[width=3.5in]{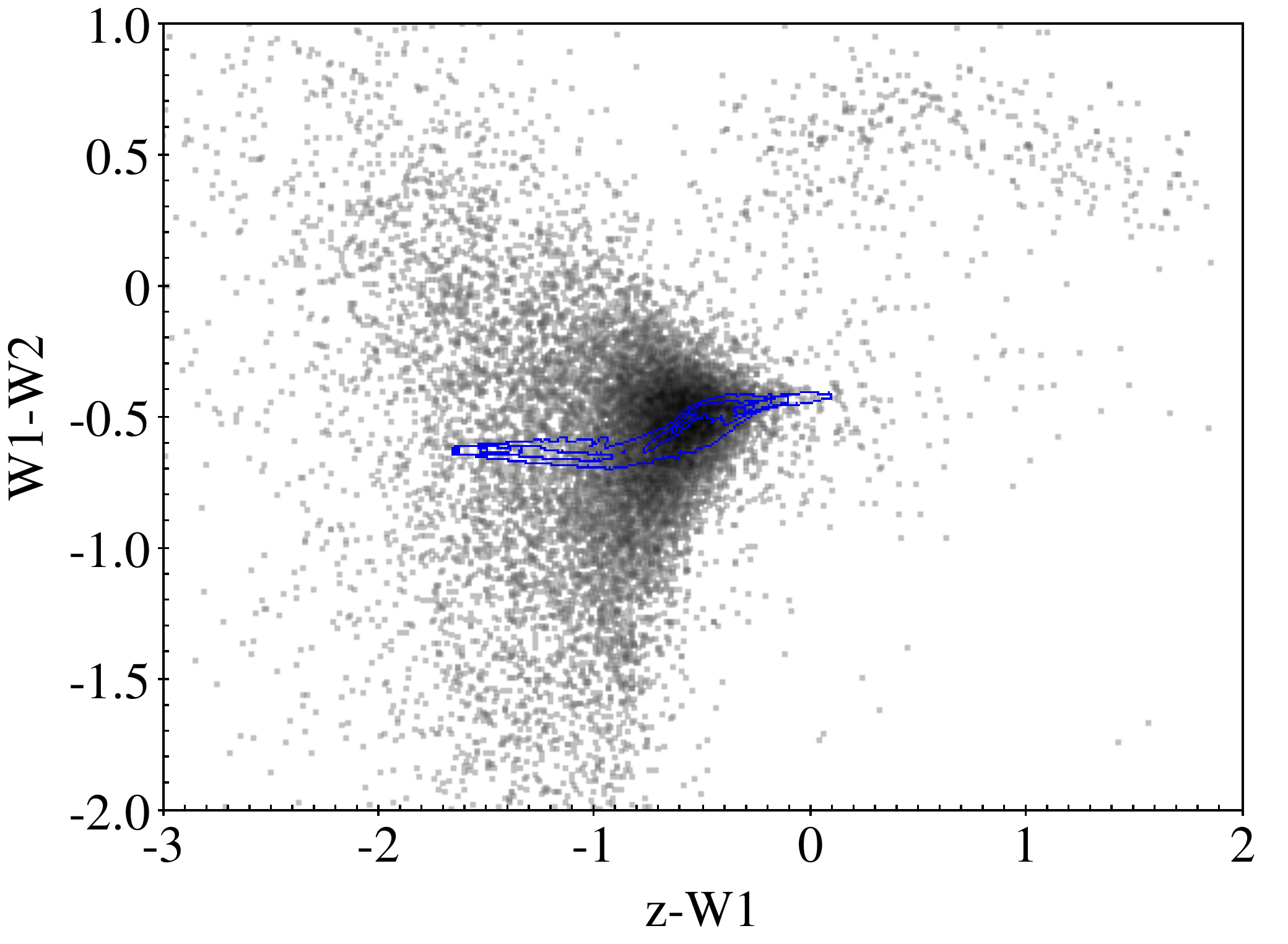}}
  \caption{\label{fig:colorcolor}Color-color diagram of $g-r$ versus $r-z$ (left panel) and $z-W1$ versus $W1-W2$ (right panel) for $19<r<20$ objects in a 20 deg$^2$ test region ($340^{\circ}<{\rm R.A.}<350^{\circ}$, $-1.0^{\circ}<{\rm Decl.}<1.0^{\circ}$). The gray dots are the DECaLS ``PSF" type objects, and the blue contours are stellar simulation colors using DECAM and WISE filters.}
\end{figure*}
Figure \ref{fig:compare} shows the spectroscopic redshift distribution compared with the Skew-t, XDQSOz, and CZR photo-$z$ distributions. The SDSS colors of $z \sim 2.2$ quasars look similar to those of $z\sim 0.8$ quasars, as the C III] and MgII lines shift into the $g$ band at $z \sim 2.2$ and $z \sim 0.8$, respectively. This is a degeneracy, if we use only the SDSS photometry to calculate the photo-$z$ of $z\sim 2.2$ and $z\sim 0.8$ quasars. We present Anderson-Darling goodness of fit tests with the R $kSamples$ package\footnote{https://cran.r-project.org/web/packages/kSamples}. Randomly choosing 1000 objects from the sample and comparing with spectroscopic redshift, the probability values of photo-$z$ from method Skew-t, XDQSOz, and CZR are 0.207, 0.102, and approximately 0, respectively. The Skew-t method performs better than other methods on this problem. A better result requires more photometric data, such as infrared photometry.

\subsection{Photo-$z$ Results Using Optical and Mid-Infrared Photometry} \label{subsec:photoz_infrared}
Adding the infrared photometric data will make the photo-$z$ results more robust \citep{Wu2010, Wu2012}. Using SDSS/PS1/DECaLS optical photometry with WISE W1 and W2 mid-infrared photometry, the $\sigma(\Delta z)$ values are reduced to 0.15, 0.18, and 0.17, respectively (Table \ref{tab:photoz_infrared}). The photo-$z$ accuracy $R_{0.1}$ is enhanced to 87.0\%, 89.1\%, and 72.4\%, respectively. As WISE data are shallower than optical data, using WISE photometry will reduce the number of quasars detected. Figure \ref{fig:photoz} shows the photo-$z$ versus spectroscopic redshift for SDSS, SDSS-WISE, PS1-WISE, and DECalS-WISE, respectively. Figure \ref{fig:offset} shows their $\Delta z$ distributions. Figure \ref{fig:raito} shows the photo-$z$ accuracy $R_{0.1}$ as a function of redshift (left panel) and $r$-band magnitude (right panel), respectively. The degeneracy problem between $z \sim 2.2$ and $z \sim 0.8$ is alleviated with the inclusion of mid-infrared data. PS1 and DECaLS do not have $u$-band data, so the photo-$z$ results are less accurate than those for SDSS-WISE. At $z>3.4$, the Lyman limit moves out of the $u$ band, and then the $u$ band photometry is not important for the photo-$z$ regression any more. Optical and mid-infrared data are sufficient for photo-$z$ regression at $3.4<z<5$. The photo-$z$ results derived without prior probability in Equation (\ref{eq:10}) are also listed in Table \ref{tab:photoz_infrared}. The $\sigma(\Delta z)$ values increase 0.1-0.5, and the photo-$z$ accuracy decreases 0.1\%-2.1\% without prior probability.

\section{The Classification method} \label{sec:classification}
\subsection{Stellar Simulation} \label{subsec:simulation}
Stars are the main contaminants for quasar candidate selection. \citet{Fan1999} simulated the SDSS colors of Galactic stars and showed that the simulated colors of stars are in good agreement with observations in the SDSS. He proved that stellar simulation can be used as a tool to separate stars and quasars. \citet{Robin2003} built a synthesis model of stellar populations (the Besan\c{c}on Galaxy Model\footnote{http://model.obs-besancon.fr}) consisting of stars from 4 populations including thin disk, thick disk, spheroid, and bulge. Each population is described by a star formation rate history, an Initial Mass Function and an age (or age-range) (see Table 1 in \citet{Robin2003}). Density laws for the thin disk are constrained self-consistently by the potential in the Boltzmann equation (see Table 3 in \citet{Robin2003}). \citet{Sharma2011} presented a code called GALAXIA creating a synthetic survey of the Milky Way based on parameters in the Besan\c{c}on model. We built up the stellar color distribution with a simulation of the Milky Way using the GALAXIA code within 30 kpc, and updated the Padova isochrones to the PARSEC isochrones \citep[PARSEC v1.2S\footnote{http://stev.oapd.inaf.it/cmd};][]{Bressan2012, Chen2014, Chen2015, Tang2014}.

Figure \ref{fig:hist} shows an example of a stellar simulation with DECAM filters in a 20 deg$^2$ test region ($340^{\circ}<{\rm R.A.}<350^{\circ}$, $-1.0^{\circ}<{\rm Decl.}<1.0^{\circ}$). Figure \ref{fig:hist} shows the $r$-band magnitude distribution of the DECaLS point objects (${\rm type} =``{\rm PSF}"$, black histogram) compared with the simulated stars (blue histogram). The gray dashed histogram shows ``PSF" type objects with $\Delta \chi^2>40$, where $\Delta \chi^2$ is the $\chi^2$ difference between fitting to a PSF model and a simple galaxy model. Smaller $\Delta \chi^2$ means that the object is more likely to be similar to a galaxy morphology. Point-like galaxy is possible to be identified as ``PSF" morphology but remains smaller $\Delta \chi^2$ comparing to those objects with true ``PSF" morphologies. The difference between the simulation and observation at $r \sim 24$ are mainly caused by the contaminations of point-like galaxies. At fainter magnitudes, the contamination by point-like galaxies becomes more and more prominent. At the faint end, the observation is not consistent with the simulation due to the magnitude limit of the imaging survey, which is roughly 23.9 mag in the r band for DELS. Figure \ref{fig:colorcolor} shows the $g-r$ versus $r-z$ (left panel) and $z-W1$ versus $W1-W2$ (right panel) color-color diagrams of stars with $19<r<20$ in this region with DECAM and WISE detections. The simulated colors (blue contours) trace the observed stellar locus (gray dots) well, so objects deviating from the stellar locus can be easily found.

The probability of one object to be a star is expressed as
\vspace{-0.1cm}
\begin{equation} \label{eq:12}
\vspace{-0.2cm}
  P_{\rm Star} = \sum_{i}^{N_{\rm Star}(m)}{P_{\rm Star}(\bm{f}|m)},
\end{equation}
where $N_{\rm Star}(m)$ is the number of stars with magnitude bin $\Delta m=0.1$ in 1 deg$^2$, and $P_{\rm Star}(\bm{f}|m)$ is expressed as a multivariate Gaussian distribution
\begin{equation} \label{eq:13}
  P_{\rm Star}(\bm{f}|m) = N(\bm{\mu}, \bm{\Sigma^*}),
\end{equation}
where $\mu$ comes from the relative fluxes of simulated stars. For example, in the case of using the DECaLS $grz$ and WISE W1 and W2 magnitudes, we use $f_g/f_r$, $f_z/f_r$, $f_{W1}/f_r$ and $f_{W2}/f_{r}$, as most stars do not vary significantly and the photometric error in the $r$-band is smaller than that in the W1 band. There are no errors assigned to the simulated relative fluxes, and $\bm{\Sigma^*}$ is the covariance matrix for a target with flux uncertainties $e_g, e_r, e_z, e_{W1}, e_{W2}$, namely
\begin{equation} \label{eq:14}
{\bm{\Sigma}^*} =
\left(
\begin{array}{ccccc}
\frac{e_g^2 f_r^2 + e_r^2 f_g^2}{f_r^4} & 0 & 0 \\
0 & \frac{e_z^2 f_r^2 + e_r^2 f_z^2}{f_r^4} & 0 & 0 \\
0 & 0 & \frac{e_{W1}^2 f_r^2 + e_r^2 f_{W1}^2}{f_r^4} & 0 \\
0 & 0 & 0 & \frac{e_{W2}^2 f_r^2 + e_r^2 f_{W2}^2}{f_r^4}\\
\end{array}
\right)
\end{equation}
We do not apply Galactic extinction to the observed data or the simulated relative fluxes when calculating the probability to be a star. The logarithmic likelihood of an object to be a star is defined as
\begin{equation} \label{eq:15}
L_{\rm Star} = {\rm log}(P_{\rm Star}).
\end{equation}

\begin{figure}[htbp]
\hspace*{-0.5cm}
\epsscale{1.2}
\plotone{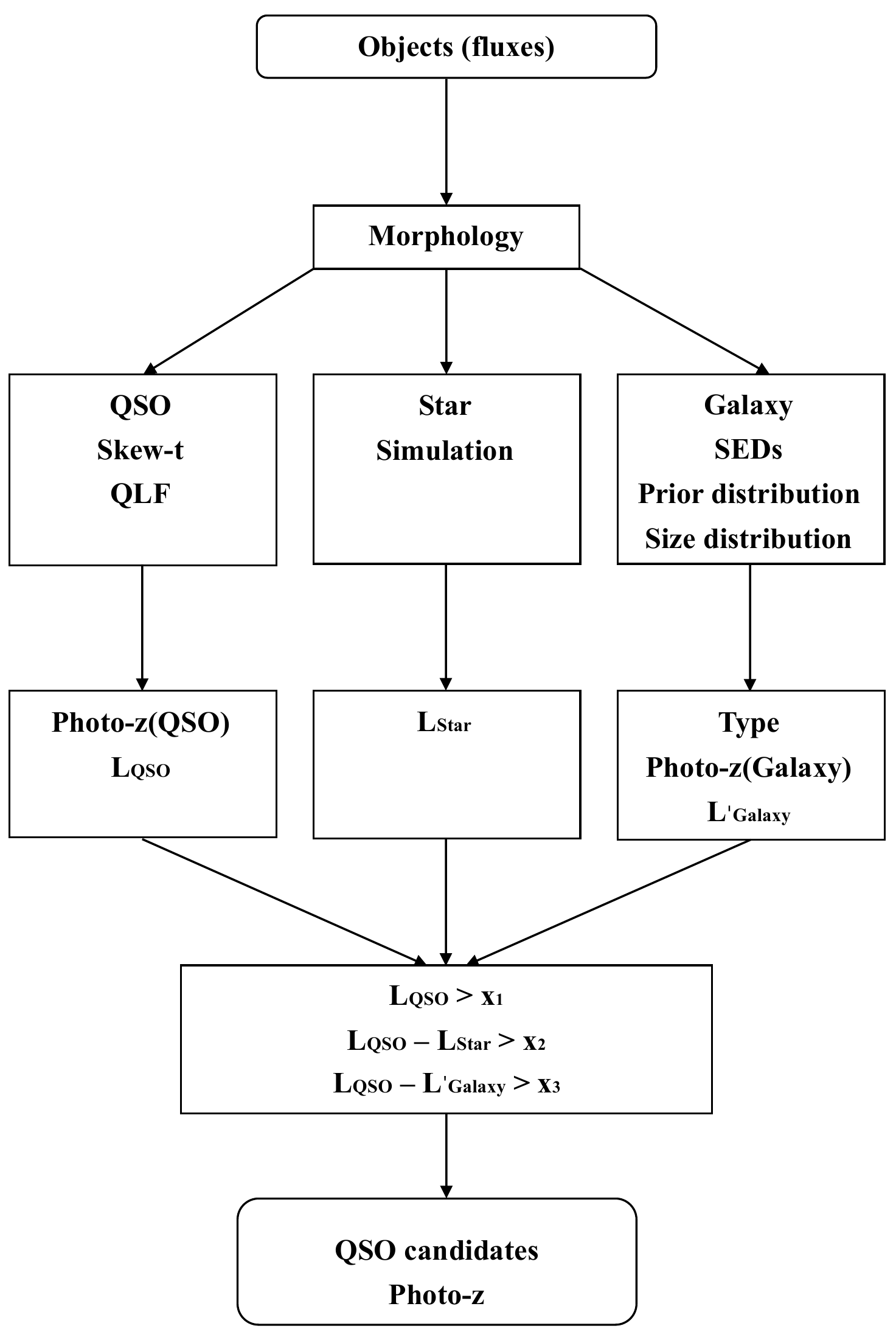}
\caption{\label{fig:flowchart}Quasar candidate selection flowchart. The classification process includes morphological selection and fitting the relative fluxes with quasar, star and galaxy fluxes. Comparing the three probabilities with Bayesian theory, we can tell whether an object is a quasar or not. The $x_1$ criterion is used to insure that the colors are not far away from the quasar colors, $x_2$ is used to establish that it is more probable to be a quasar than a star, and $x3$ to tell that it is more probable to be a quasar than a galaxy.}
\end{figure}

\begin{figure*}
  \centering
  \hspace{-0.3cm}
    \subfigure{
    \includegraphics[width=3.7in]{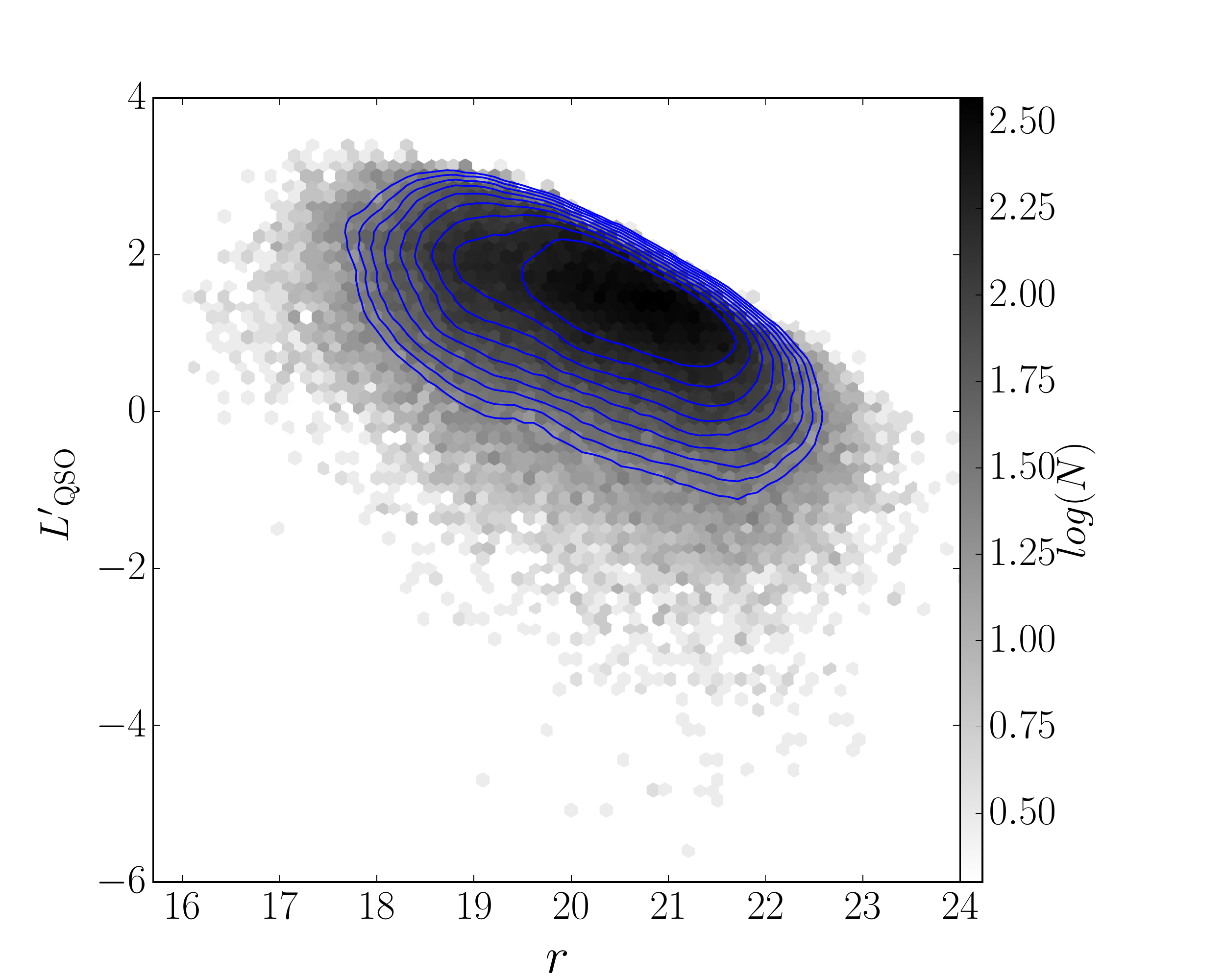}}
  \hspace{-1.0cm}
  \subfigure{
    \includegraphics[width=3.7in]{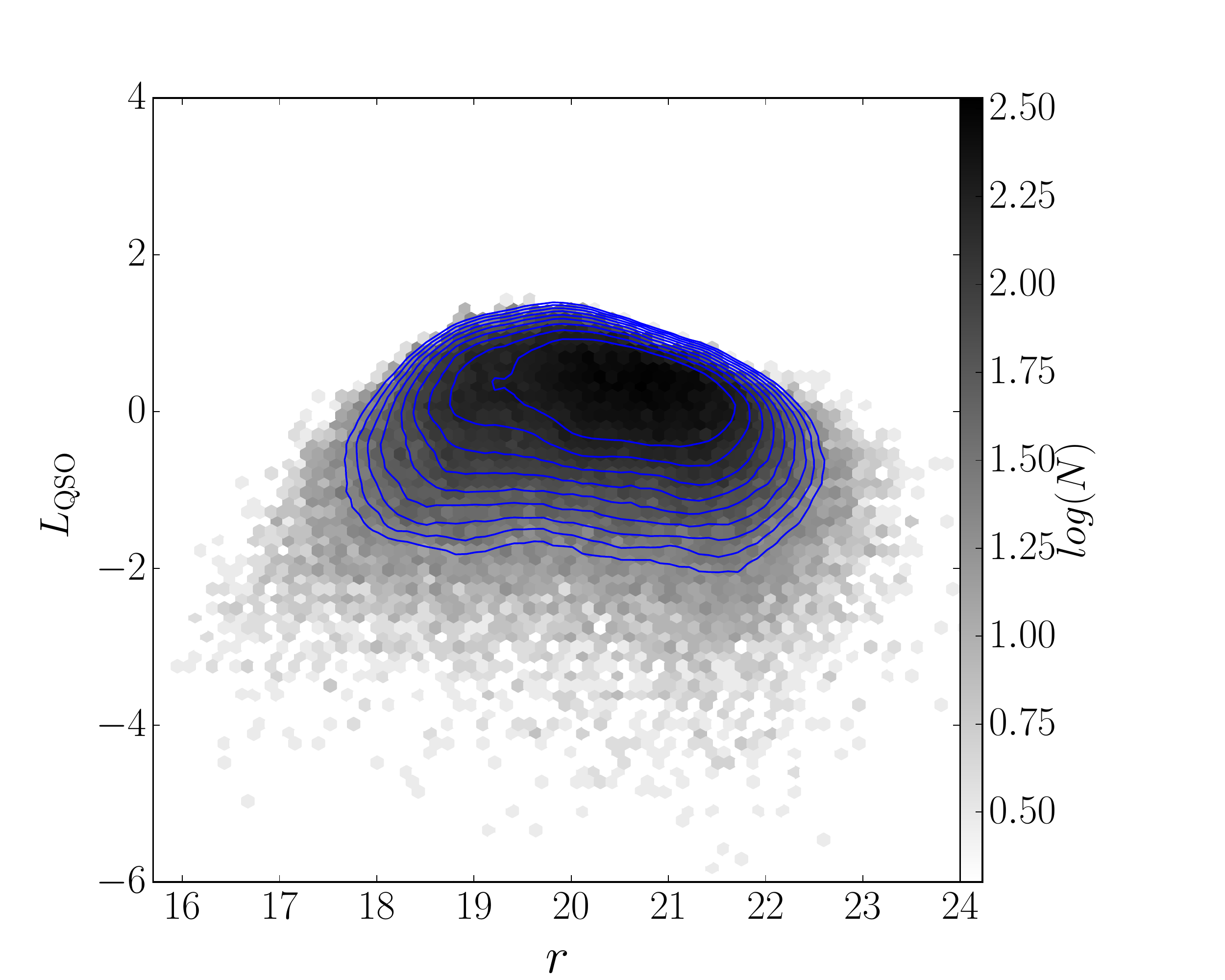}}
  \caption{\label{fig:prior}The logarithmic likelihoods of objects to be quasars from the posterior probability $L'_{\rm QSO}$ in Equation (\ref{eq:11}) (left panel), and those from a convolution of the posterior probability and the prior probability $L_{\rm QSO}$ in Equation (\ref{eq:9}) (right panel). The x-axis is $r$-band magnitude. The photometry used in these two panels are the DECaLS $g, r,$ and $z$-band photometry, and WISE W1 and W2-band photometry. Density contours are in blue. For fainter objects, the posterior probabilities decrease due to the larger photometric uncertainties. According to the QLF, a fainter object is more likely to be a quasar compared to a brighter object. $L_{\rm QSO}$ is more uniform than $L'_{\rm QSO}$ over the range of magnitudes.}
\end{figure*}

\subsection{Galaxy template fitting} \label{subsec:galaxy}
It becomes more difficult to classify a fainter object as star or galaxy by morphology, thus the contamination from point-like galaxies becomes more significant. We reduce the contamination of point-like galaxies by a galaxy template fitting procedure. For the dataset that includes WISE photometry, we use a subsample of 18 galaxy spectral energy distributions (SEDs) from \citet{Brown2014} with wavelength coverage from ultraviolet to mid-infrared. \citet{Benitez2000} presented a BPZ algorithm to estimate the photo-$z$ of galaxies. They derived the galaxy prior probability as $p(z, {\rm type}, m)=p({\rm type}|m)p(z|{\rm type}, m)$, where $p({\rm type}|m)$ is the galaxy type fraction as a function of magnitude, and $p(z|{\rm type}, m)$ is the redshift distribution of a given galaxy type and magnitude. As many imaging observations were taken under 1\arcsec seeing, galaxies with half-light radius $R_e<0.5\arcsec$ \citep{Fan1999} are treated as point-like galaxies. \citet{Shen2003} obtained the size distributions of early and late type galaxies and their dependence on luminosity from 140,000 SDSS galaxies. We derive the prior distribution $p(z, {\rm type}, m)$ of point-like galaxy for spectral types corresponding to E/S0, Sbc/Scd and Irr (parameters from Table 1 in \citet{Benitez2000}) with $R_e<0.5\arcsec$. The probability of one object being a galaxy over the whole redshift range is expressed as
\begin{equation} \label{eq:16}
  P_{\rm Galaxy} = \sum\sum_{i}\int{P_{\rm Galaxy}(z)N_{\rm Galaxy}(z, {\rm type},m)dz},
\end{equation}
where $\Sigma$ is the sum of all galaxy types, and $\Sigma_{i}$ is the sum of all galaxy SEDs for each type, and $N_{\rm Galaxy}(z, {\rm type},m)$ is the number of a certain type of galaxy at redshift z and magnitude m within $\Delta z = 0.05$ and $\Delta m = 0.1$ per $deg^2$. $P_{\rm Galaxy}(z)$ is expressed as a multivariate Gaussian distribution
\begin{equation} \label{eq:17}
  P_{\rm Galaxy}(z) = N(\bm{\mu}(z), \bm{\Sigma}^*) ,
\end{equation}
where $\mu(z)$ comes from the relative fluxes of galaxy templates as a function of redshift, and $\Sigma^*$ is described in Equation (\ref{eq:14}).
$N_{\rm Galaxy}(z, {\rm type},m)$ is derived from the prior distribution $p(z, {\rm type}, m)$ using a scale factor c, and defining $P'_{\rm Galaxy}$ as
\begin{equation} \label{eq:18}
  P'_{\rm Galaxy} = \sum\sum_{i}\int{P_{\rm Galaxy}(z)p(z, {\rm type},m)dz},
\end{equation}
The logarithmic likelihood of an object being a galaxy is defined as
\begin{equation} \label{eq:19}
L_{\rm galaxy} = {\rm log}(P_{\rm galaxy}) = {\rm log(c)} + L'_{\rm galaxy},
\end{equation}
where
\begin{equation} \label{eq:20}
L'_{\rm galaxy} = {\rm log}(P'_{\rm galaxy}).
\end{equation}

\subsection{Quasar Candidate Selection Flowchart} \label{subsec:flowchart}
Quasars can be selected based on Bayesian probabilities \citep[e.g.,][]{Richards2009b, Richards2015, Kirkpatrick2011}. Considering a point-like object that is likely to be a quasar, star or galaxy, the Bayesian probability of being a quasar is expressed as
\begin{equation} \label{eq:21}
P({\rm QSO}) = \frac{P_{\rm QSO}}{P_{\rm QSO} + P_{\rm Star} + P_{\rm Galaxy}}
\end{equation}
where $P_{\rm QSO}$, $P_{\rm Star}$ and $P_{\rm Galaxy}$ are expressed in Equations (\ref{eq:7}), (\ref{eq:12}) and (\ref{eq:17}), respectively. A Bayesian probability criterion is usually defined as $P({\rm QSO})>x$, namely
\begin{equation} \label{eq:22}
\frac{P_{\rm Star}}{P_{\rm QSO}} + \frac{P_{\rm Galaxy}}{P_{\rm QSO}} < \frac{1}{x} - 1.
\end{equation}
Here we suggest three Bayesian probability criteria as
\begin{equation} \label{eq:23}
L_{\rm QSO} > x_1,
\end{equation}
\begin{equation} \label{eq:24}
L_{\rm QSO} - L_{\rm Star} > x_2,
\end{equation}
\begin{equation} \label{eq:25}
L_{\rm QSO} - L'_{\rm Galaxy} > x_3,
\end{equation}
which correspond to
\begin{equation} \label{eq:26}
P_{\rm QSO} > 10^{x_1}
\end{equation}
\begin{equation} \label{eq:27}
\frac{P_{\rm Star}}{P_{\rm QSO}} < 10^{x_2}
\end{equation}
\begin{equation} \label{eq:28}
\frac{P_{\rm Galaxy}}{P_{\rm QSO}} < c10^{x_3}
\end{equation}
These criteria mean that (1) the object has relative fluxes similar to quasars, (2) the object is more likely to be a quasar than a star, (3) the object is more likely to be a quasar than a galaxy. The quasar candidate selection flowchart is shown in Figure \ref{fig:flowchart}. For a given object, we measure its relative fluxes and magnitudes, and then apply a morphology criterion that most quasars are point-like objects. Then we calculate the probability of the object being (1) a quasar, with a prior probability derived from the QLF, and a posterior probability modeled with a multivariate Skew-t distribution as a function of magnitude and redshift; (2) a star, with a prior probability from number counts and distribution of stellar parameters from a Milky Way synthetic simulation, and a posterior distribution modeled by a multivariate Gaussian distribution with relative fluxes from the Padova isochrones; (3) a galaxy, with a prior probability from the BPZ prior distribution for point-like galaxies, and a posterior probability modeled by a multivariate Gaussian distribution with relative fluxes from galaxy templates. We obtain quasar candidates, as well as photo-$z$s, with the three Bayesian probability criteria in Equations (\ref{eq:23}), (\ref{eq:24}), (\ref{eq:25}).

For fainter objects, the posterior probabilities decrease due to larger photometric uncertainties. The left panel in Figure \ref{fig:prior} shows the logarithmic likelihoods of the objects to be quasars integrated from the posterior probability ($L'_{\rm QSO}$ in Equation (\ref{eq:11})). According to the QLF, a fainter object is more likely to be a quasar compared to a brighter object. The right panel in Figure \ref{fig:prior} shows the logarithmic likelihoods integrated from posterior probabilities and prior probabilities ($L_{\rm QSO}$ in Equation (\ref{eq:9})). $L_{\rm QSO}$ is more uniform than $L'_{\rm QSO}$ over the range of magnitudes. A criterion of a $L_{\rm QSO}$ cut is more reasonable than a simple $\chi^2$ cut or a probability cut without considering photometric uncertainties. As a consequence, the selection completeness will be affected by the prior distribution from the QLF. For example, if the bright end of the QLF is underestimated, some bright quasars with colors deviating from bright normal quasars may be missed. 

\begin{figure*}
\centering
\hspace{0cm}
\epsscale{1.1}
\plotone{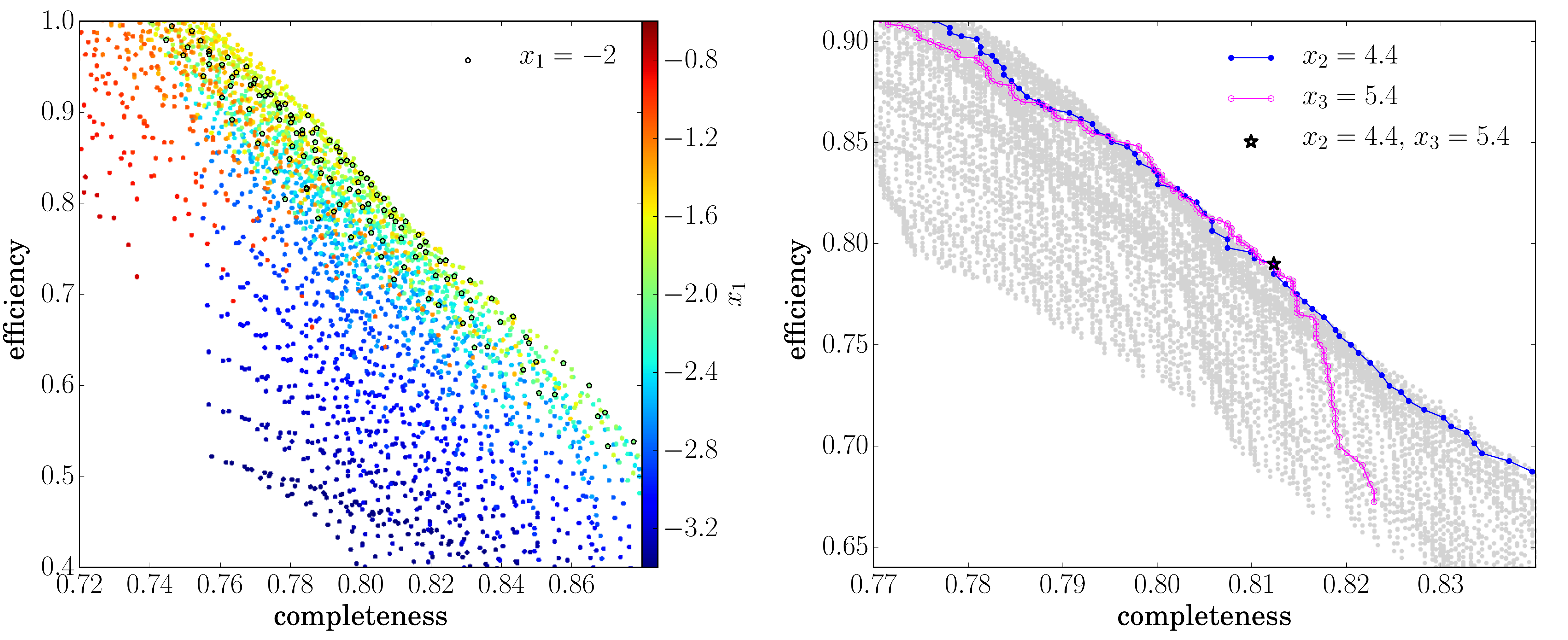}
\caption{\label{fig:roc} Trade-off between completeness and efficiency in the classfication procedure. Left panel: the dots are results with $x_1 \in (-3.5, -0.5)$, $x_2 \in (0, 15)$ and $x_3 \in (0, 10)$, and the color map shows the value of $x_1$. Black pentagons mark $x_1 = -2$, which are obviously located at the edge of the PR diagram. Right panel: results with $x_1 = -2.0$, $x_2 \in (0, 15)$ and $x_3 \in (0, 10)$. The blue dot-line denotes $x_2 = 4.4$, and is located at the edge of the curves for a wide range of $x_3$ values. The magenta open dot-line denotes $x_3 = 5.4$, and is located at the edge of the curves when the completeness is in the range of $\sim 79\%$ to $\sim 82\%$. The black unfilled star marks the point $x_1 = -2$, $x_2 = 4.4$, and $x_3 = 5.4$.}
\end{figure*}

\begin{figure*}
\centering
\hspace{0cm}
\epsscale{1.1}
\plotone{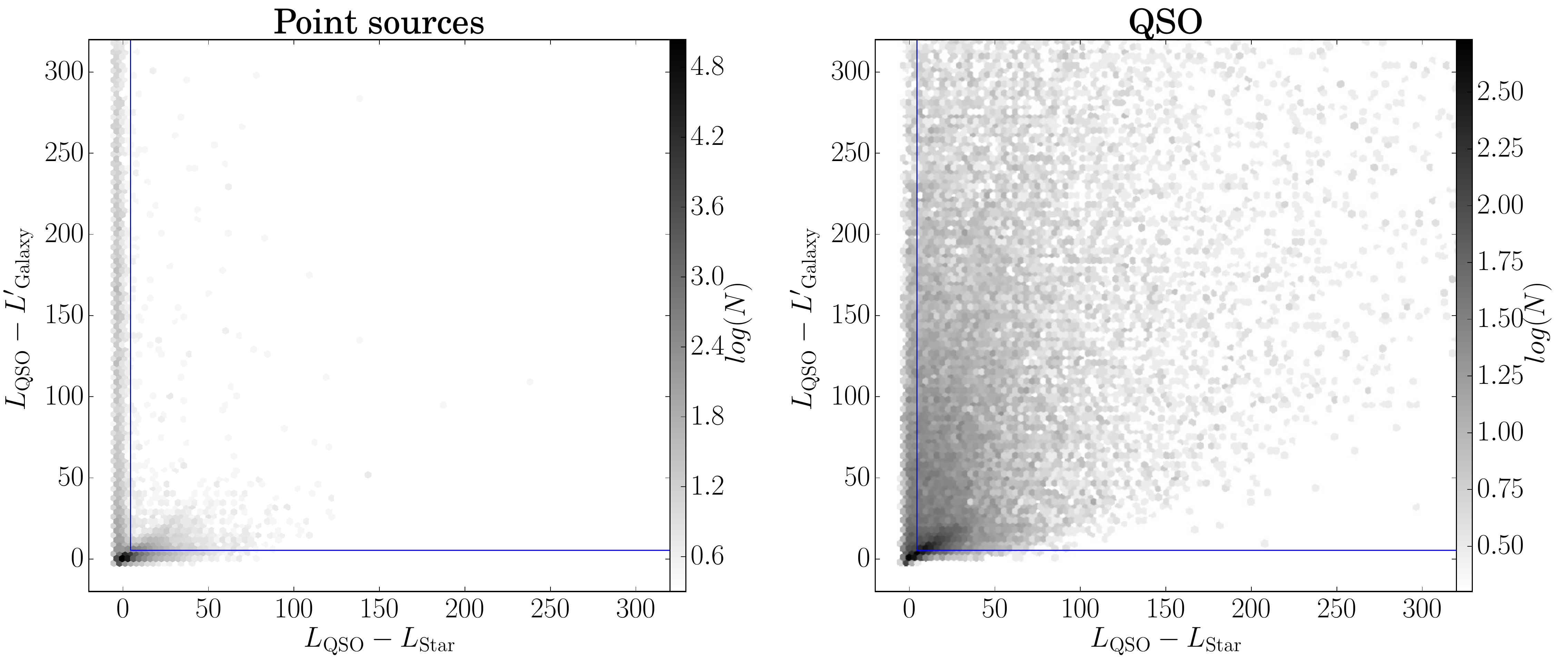}
\caption{\label{fig:ll} The $L_{\rm QSO}-L'_{\rm Galaxy}$ versus $L_{\rm QSO}-L_{\rm Star}$ diagram. The gray scale hexagon show the density of point sources (left panel) and quasars (right panel). The blue vertical line is the cut of $L_{\rm QSO}-L_{\rm Star} = 4.4$, and the blue horizontal line is the cut of $L_{\rm QSO}-L'_{\rm Galaxy} = 5.4$. Quasars span a much larger space in the $L_{\rm QSO}-L'_{\rm Galaxy}$ versus $L_{\rm QSO}-L_{\rm Star}$ diagram, and 11\% of known quasars with ``PSF" type and $L_{\rm QSO} >-2$ are located below these two cuts. Meanwhile, 85\% of point objects with $L_{\rm QSO}>-2$ are excluded by these two cuts.}
\end{figure*}

\section{Discussion} \label{sec:discussion}
\subsection{Quasar Candidate Selection using DECaLS and WISE photometry} \label{subsec:selection}
We test the quasar candidate selection algorithm described in Section \ref{subsec:flowchart} with the DECaLS $g$, $r$, $z$ and WISE W1 and W2 photometry. There is a 15 deg$^2$ region ($36^{\circ}<{\rm R.A.}<42^{\circ}$ and $-1.25^{\circ}<{\rm Decl.}<1.25^{\circ}$) in S82 with spectroscopically identified quasars as dense as 167 per deg$^2$. We exclude quasars in this region from the quasar training sample, and model the quasar relative fluxes posterior distribution using the method in Section \ref{subsec:skewt}. For selection criteria, there is a trade-off between completeness and efficiency. We define the completeness as the completeness of selecting the spectroscopically identified quasars at $r<23$ mag and redshift $z<5.4$. The efficiency is defined as
\begin{equation} \label{eq:29}
{\rm efficiency} = \frac{N_{\rm QLF}(r) * {\rm completeness}(r)}{N_{{\rm photo}-z}(r)},
\end{equation}
where $N_{\rm QLF}(r)$ is calculated from the QLF \citep{NPD2016} (PLE+LEDE model). The completeness also includes $\sim 5\%$ incompleteness from the ``PSF" morphological criterion. It is worth noting that the completeness is probably overestimated, because the spectroscopic sample is not complete \citep{Ross2012, Ross2013} at $r<23$ mag even in this dense S82 region. Therefore, the efficiency might be also overestimated. Figure \ref{fig:roc} shows the efficiency vesus completeness, the Precision-Recall (PR) diagram \citep{Davis2006}, with parameters $x_1 \in (-3.5, -0.5)$, $x_2 \in (0, 15)$ and $x_3 \in (0, 10)$. These parameter ranges are large enough to cover a wide range of the completeness and efficiency space. We determine $x_1$, $x_2$, and $x_3$ sequentially. First, for the above given ranges of $x_2$ and $x_3$, the best value $x_1$ is --2. The black pentagons mark $x_1 = -2$, which is located at the edge of the PR diagram, with relatively larger efficency with the same completeness. With the criterion $L_{\rm QSO} >-2$, 97\% of the known quasars (``PSF" type) are selected, and 87\% of the point sources are excluded. For the fixed $x_1=-2$ and the $x_3$ range given above, the best $x_2$ value is 4.4. The blue dot-line denotes $x_2 = 4.4$, which is located at the edge of the PR with a wide range of $x_3$ values. Finally, with the best values of $x_1$ and $x_2$ determined above, we find the best $x_3$ to be 5.4. The magenta open dot-line shows where $x_3 = 5.4$, which is located at the edge when the completeness is in the range $\sim 79\%$ to $\sim 82\%$. The black star marks $x_2 = 4.4$ and $x_3 = 5.4$. Figure \ref{fig:ll} shows the $L_{\rm QSO}-L'_{\rm Galaxy}$ versus $L_{\rm QSO}-L_{\rm Star}$ diagram. Quasars span a much larger space in the $L_{\rm QSO}-L'_{\rm Galaxy}$ versus $L_{\rm QSO}-L_{\rm Star}$ diagram, and 11\% of the known quasars with ``PSF" type and $L_{\rm QSO} >-2$ are located below these two cuts. Meanwhile, 85\% of the point objects with $L_{\rm QSO}>-2$ are excluded by these two cuts. Larger cuts will result in lower selection completeness and higher efficiency. For example, specifically a $L_{\rm QSO}-L_{\rm Star} > 10.0$ cut excludes 86\% of the point objects with $L_{\rm QSO}>-2$, but meanwhile causes 12\% more selection incompleteness at $z\sim 2.8$. Therefore, we suggest the criteria for quasar selection when using the DECaLS $g$, $r$, $z$, and WISE W1 and W2 to be
\begin{equation} \label{eq:30}
{\rm type} = ``{\rm PSF}",
\end{equation}
\begin{equation} \label{eq:31}
{\rm flux}(r)>0,
\end{equation}
\begin{equation} \label{eq:32}
L_{\rm QSO}>-2,
\end{equation}
\begin{equation} \label{eq:33}
L_{\rm QSO} - L_{\rm Star}>4.4,
\end{equation}
\begin{equation} \label{eq:34}
L_{\rm QSO} - L'_{\rm Galaxy}>5.4.
\end{equation}

\begin{figure*}[htbp]
\hspace*{-0.5cm}
\epsscale{1.2}
\plotone{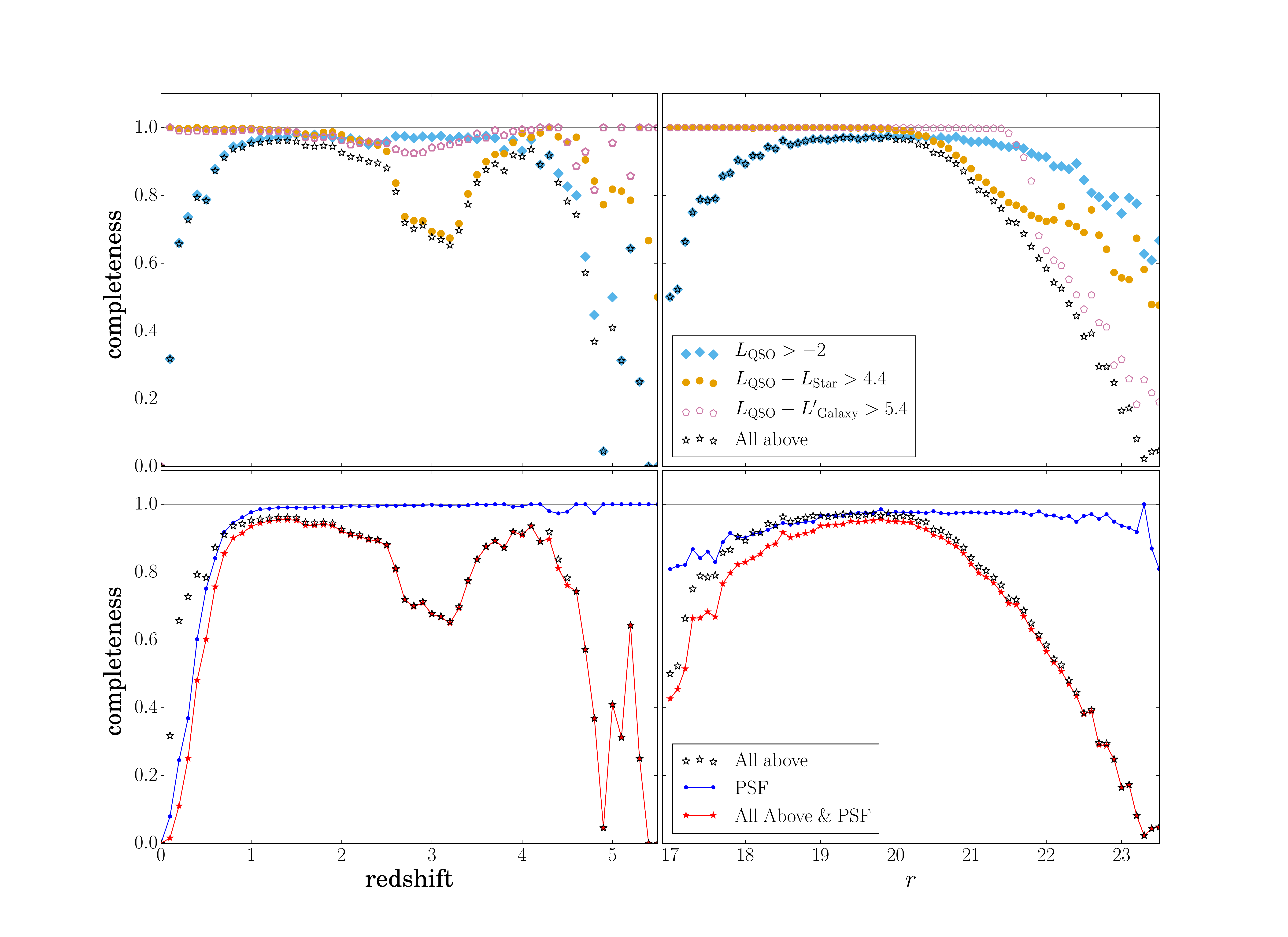}
\caption{\label{fig:completeness}The completeness of the classification method used for the spectroscopic quasar sample as a function of redshift (left panels) and $r$-band magnitude (right panels). The top panels show the completeness results applying the logarithmic likelihood criteria, including criterion $L_{\rm QSO}>-2$ in Equation (\ref{eq:32}) (blue diamonds), criterion $L_{\rm QSO} - L_{\rm Star}>4.4$ in Equation (\ref{eq:33}) (orange dots), criterion $L_{\rm QSO} - L'_{\rm Galaxy}>5.4$ in Equation (\ref{eq:34}) (magenta open pentagons), and criteria all above in Equations (\ref{eq:32})-(\ref{eq:34}) (black open stars). The bottom panels show the completeness results applying the ``PSF" morphology criterion in Equation (\ref{eq:30}) (blue dot-line), and criteria combined all logarithmic likelihood criteria and ``PSF" morphology criterion in Equations (\ref{eq:30})-(\ref{eq:34}) (red star-line). The incompleteness for $z<1$ is probably caused by quasar variability, non-PSF morphology, and host galaxy contamination. $z\sim 2.8$ quasars are close to the stellar locus, and the completeness of $z\sim2.8$ decreases to $\sim 70\%$. The completeness decreases to lower than 50\% at $r>22.3$ as the WISE data are shallower than the DECaLS data.}
\end{figure*}

\begin{figure*}
  \centering
  \hspace{-0.5cm}
    \subfigure{
    \includegraphics[width=3.7in]{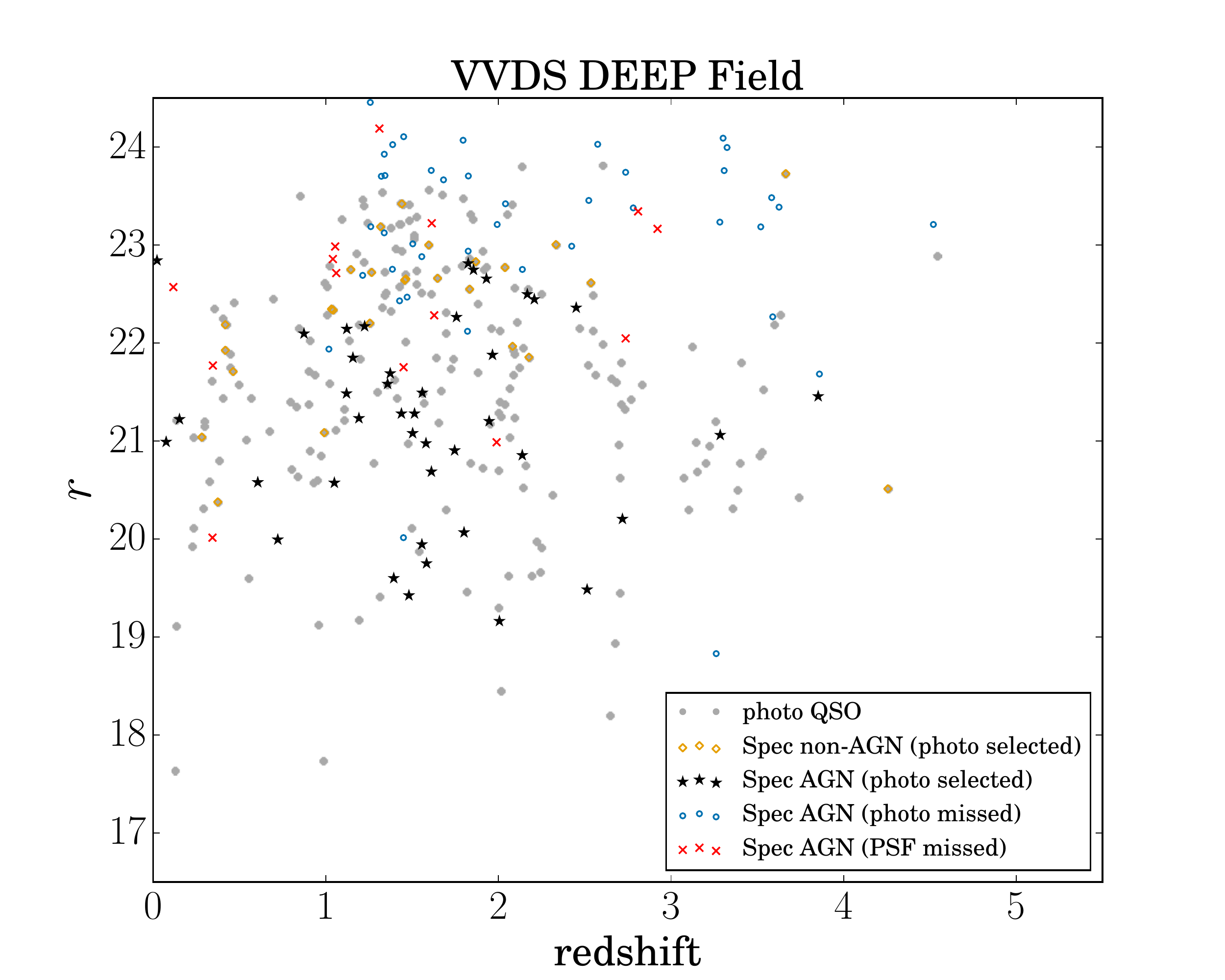}}
  \hspace{-1.2cm}
  \subfigure{
    \includegraphics[width=3.7in]{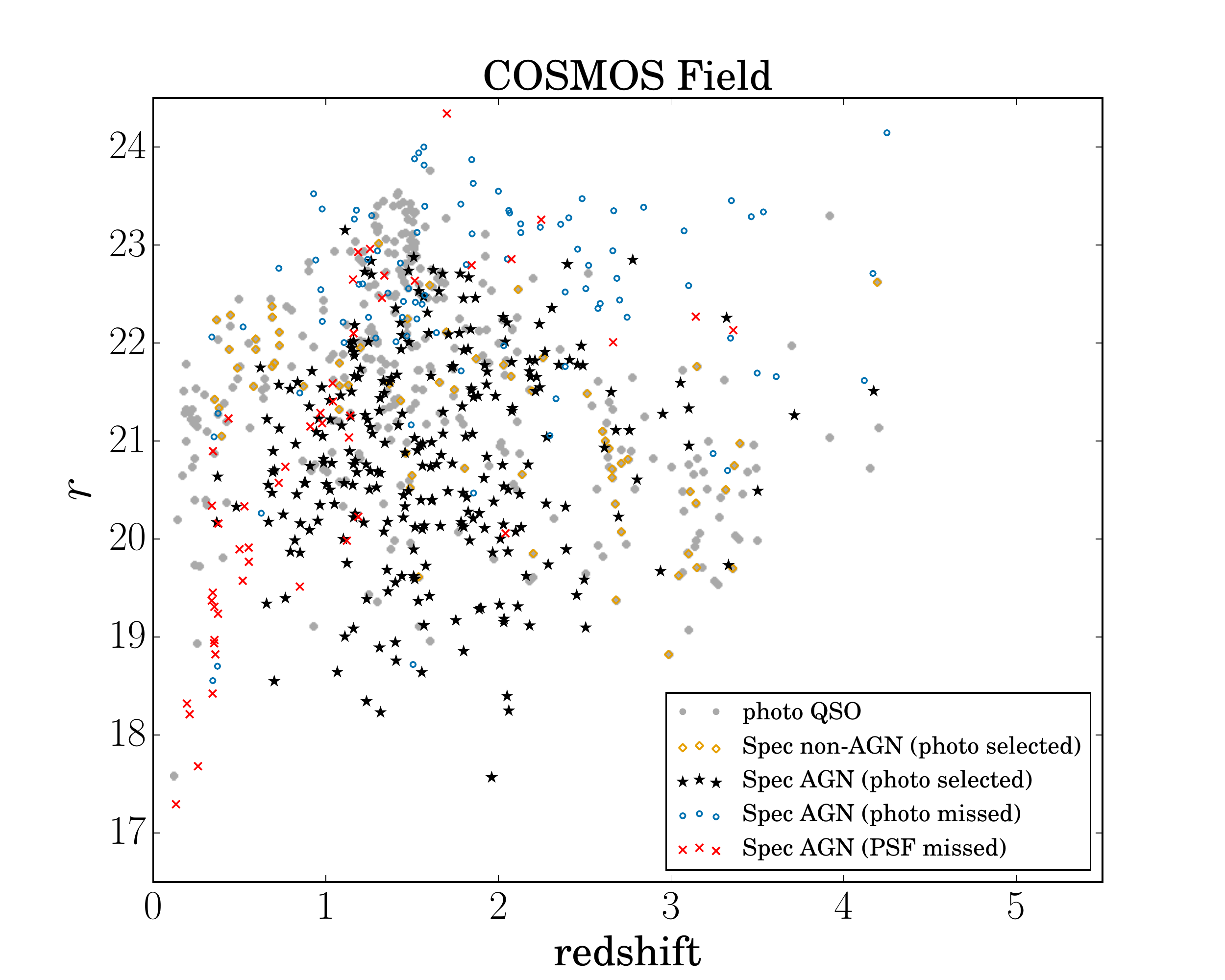}}
  \caption{\label{fig:deep}Objects in fields with deep spectroscopic surveys, in the VVDS deep field (left panel) and the COSMOS field (right panel). The y-axis is $r$-band magnitude. The black stars are spectroscopically identified AGNs that are selected by our classification method, and the blue open circles are AGNs missed by our method. Most missed AGNs are fainter than 23 mag. The gray dots are photo-$z$ selected quasars without spectra. The red crosses are AGNs missed by our method because their morphology types are not ``PSF". The orange diamonds show non-AGN objects selected by our method, and they mainly show up at $z<1$ or at the faint end.} 
\vspace{0.5cm}
\end{figure*}
\begin{table*}
\small
\begin{center}
\tablewidth{1pt}
\caption{Quasar Candidate Selection Test in Spectroscopic Surveys} \label{tab:deep}
\begin{threeparttable}
\begin{tabular}{lcrcccccc}
\tableline\tableline
${\rm Field}$ & ${\rm Area}$ & Spec\tnote{a} & photo\tnote{b} & QSO\tnote{c} & Star\tnote{d} & Galaxy\tnote{e} & completeness\tnote{f} & efficiency\tnote{g} \\
& $/deg^2$ & QSO & QSO & selected & selected & selected & spec & spec\\
\tableline
VVDS (deep)  & 0.7 & 70 & 273 & 53 & 0 &  17 & 71\% & 76\% \\
COSMOS & 2.1 & 156 & 255 & 119 & 1 & 28 & 76\% & 77\% \\
S82 (2.5h) & 15 & 153 & 271 & 129 & 8 & 3  & 84\% & 91\%\\
S82 (22.7h-3h) & 162.5 & 71 & 251 & 64 & 2 & 1 & 90\% & 95\%\\
\tableline
\end{tabular}
\tablecomments{Collumns in the table are as follows ($r<22.5$),}
\begin{tablenotes}
        \item[a] spectroscopically identified quasar number per $deg^2$.
        \item[b] photometrically selected quasar number per $deg^2$.
        \item[c/d/e] photometric method selected objects that are spectroscopically identified as quasars/stars/galaxies per $deg^2$.
        \item[f] The completeness is calculated from the spectroscopically identified quasars at $r<22.5$.
        \item[g] The efficiency is calculated as the ratio of photometric method selected, spectroscopically identified quasars from all spectroscopically identified objects (quasars, stars, and galaxies) at $r<22.5$.
    \end{tablenotes}
   \end{threeparttable}
   \end{center}
\end{table*}

With the criteria in Equations (\ref{eq:30})-(\ref{eq:34}), the selection completeness of spectroscopically identified quasars in the dense region is 81\%. For 98,450 quasars with DECaLS photometry, we recover 84,639 quasars (86\%). Figure \ref{fig:completeness} shows the completeness for the spectroscopically identified quasar sample as a function of redshift (left panel) and $r$-band magnitude (right panel). In the top panels, the blue diamonds show the completeness after applying the criteria in Equation (\ref{eq:32}). The completeness decreases when the redshift is less than 1, and one possible reason is the uncertainties from variability, because the DECaLS images and WISE images were not taken simultaneously. The incompleteness at $z>4.5$ is mainly caused by the limited number of high redshift quasars and larger photometric uncertainty in the $r$ band. Better Bayesian probability selection for $z>4.5$ quasars is potentially possible if we use simulated quasar fluxes \citep{McGreer2013}, relative fluxes divided by the $z$, $y$ or $J$-band flux, and the QLF at high redshift. The orange dots show the completeness using the criterion in Equation (\ref{eq:33}), the completeness at $z \sim 2.8$ decreases as quasars move close to the stellar locus \citep[e.g. ][]{Fan1999}. This criterion also causes an incompleteness at $r>21$ mag. The magenta open pentagons show the completeness with the criterion in Equation (\ref{eq:34}). The completeness decreases rapidly at $r>21.5$ mag, as the WISE photometric uncertainties increase dramatically. The black open stars show the completeness when applying all the criteria in Equations (\ref{eq:32})-(\ref{eq:34}). The blue dot-line in Figure \ref{fig:completeness} (bottom panels) shows the completeness applying the ``PSF" morphology criterion as a function of redshift (bottom left panel) and magnitude (bottom right panel). The completeness with the morphology criterion decreases rapidly as redshift decrease at $z<1$. The fraction of known quasars satisfying the morphology criterion decreases from 92\% with redshift at $0.5<z<1$ to 53\% at $z<0.5$. This criterion also causes an incompleteness at the bright magnitude end. As the $r$-band magnitude goes fainter than 22.7, the completeness with the morphology criterion begins to decrease. The fraction decrease from 96\% at $22.5<r<23$ to 93\% at $23<r<23.5$. The solid red stars show the completeness when applying the three criteria above in Equations (\ref{eq:32})-(\ref{eq:34}) and the ``PSF" morphology criterion in Equation (\ref{eq:30}). Because the resolutions of the WISE images and the DECaLS images are different, the extended morphology introduces high photometric uncertainties at $z<1$. Furthermore, light from host galaxies also contaminate the colors of quasars at $z<1$.

The number counts of stars vary in different locations in the sky, but such variations have little effect on quasar selection completeness and efficiency if we consider only relatively high Galactic latitude $|b|\geq 30^{\circ}$. We ran a simulation for an area of 20 deg$^2$ at $b=80^{\circ}$, and the star counts are reduced by 30\% compared to the number at $b = -50^{\circ}$. We performed tests with the relative fluxes of the simulated stars and the number counts at $b=80^{\circ}$, and applied the criteria in Equations (\ref{eq:30})-(\ref{eq:34}). The selection completeness increases by 0.05\%. A test in the 15 deg$^2$ S82 region shows that the efficiency differs by 0.22\% at $r<22.5$mag. The star number counts do not strongly affect the selection results. In this work, we use only the stellar simulation from a 20 deg$^2$ region in S82 described in Section \ref{subsec:simulation} (Galactic latitude $b\sim -50^{\circ}$).

\begin{figure}[htbp]
\hspace*{-0.5cm}
\epsscale{1.2}
\plotone{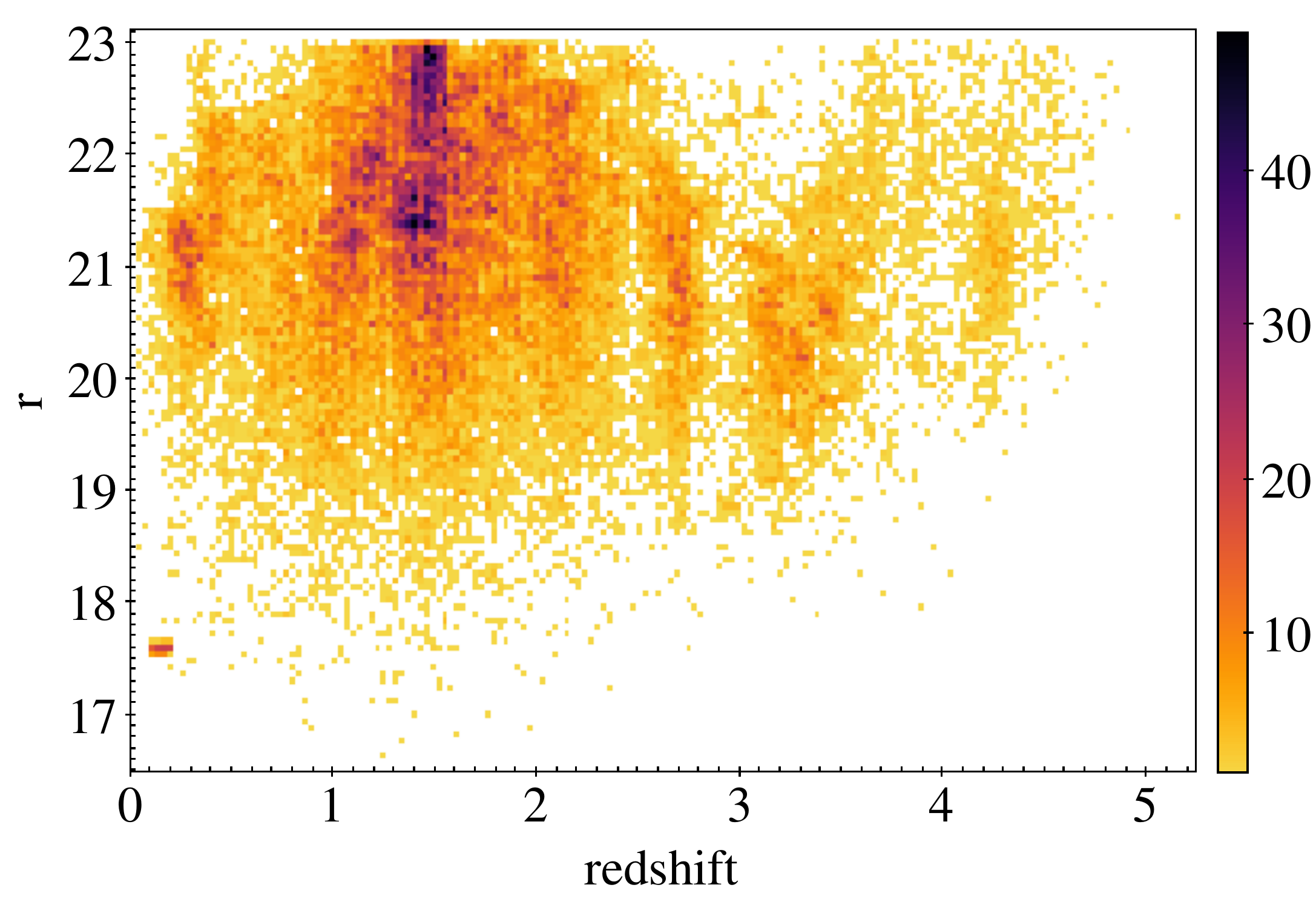}
\caption{\label{fig:dist}The $r$-band magnitude versus redshift distribution of photometric quasars in the S82 catalog. The colorbar shows the number density.}
\end{figure}

\subsection{Test Quasar Candidate Selection in Deep Fields} \label{subsec:deep}

For the dense quasar region in S82 (R.A.$\sim$2.5h), we recover 84\% of the quasars at $r<22.5$ mag (129 of 153 per $deg^2$). Only 8 stars per $deg^2$ at $r<22.5$ with spectra in the SDSS DR13 catalog\footnote{https://data.sdss.org/sas/dr13/sdss/spectro/redux/specObj-dr13.fits} are selected, and 3 galaxies per $deg^2$ at $r<22.5$ with spectra in the SDSS DR13 catalog are selected. We test the quasar selection method using the DECaLS $g$, $r$, $z$ and WISE W1, W2 photometry in some fields with deeper spectroscopic surveys. There are 104 AGNs, and 9789 galaxies in one of the VVDS deep fields \citep[$``{\rm vvds\_spF02}"$,][]{Gavignaud2006}. In the COSMOS field, there are 409 spectroscopically identified AGNs \citep{Prescott2006, Trump2009, Lilly2009}, and in the SDSS DR7\&DR12. We recover 71\% and 76\% of $r<22.5$ AGNs in the VVDS deep field and COSMOS field, respectively. More detailed results of the quasar candidate selection in some deep spectroscopic surveys at $r<22.5$ are listed in Table \ref{tab:deep}. Figure \ref{fig:deep} shows quasar selection in the VVDS deep field (left panel) and the COSMOS field (right panel). The AGNs missed are mostly because of the morphology criterion or magnitude fainter than 23 mag. The deep survey results confirm that our method performs well in quasar candidate selection.

\begin{figure*}
  \centering
  \hspace{0cm}
    \subfigure{
    \includegraphics[width=3.5in]{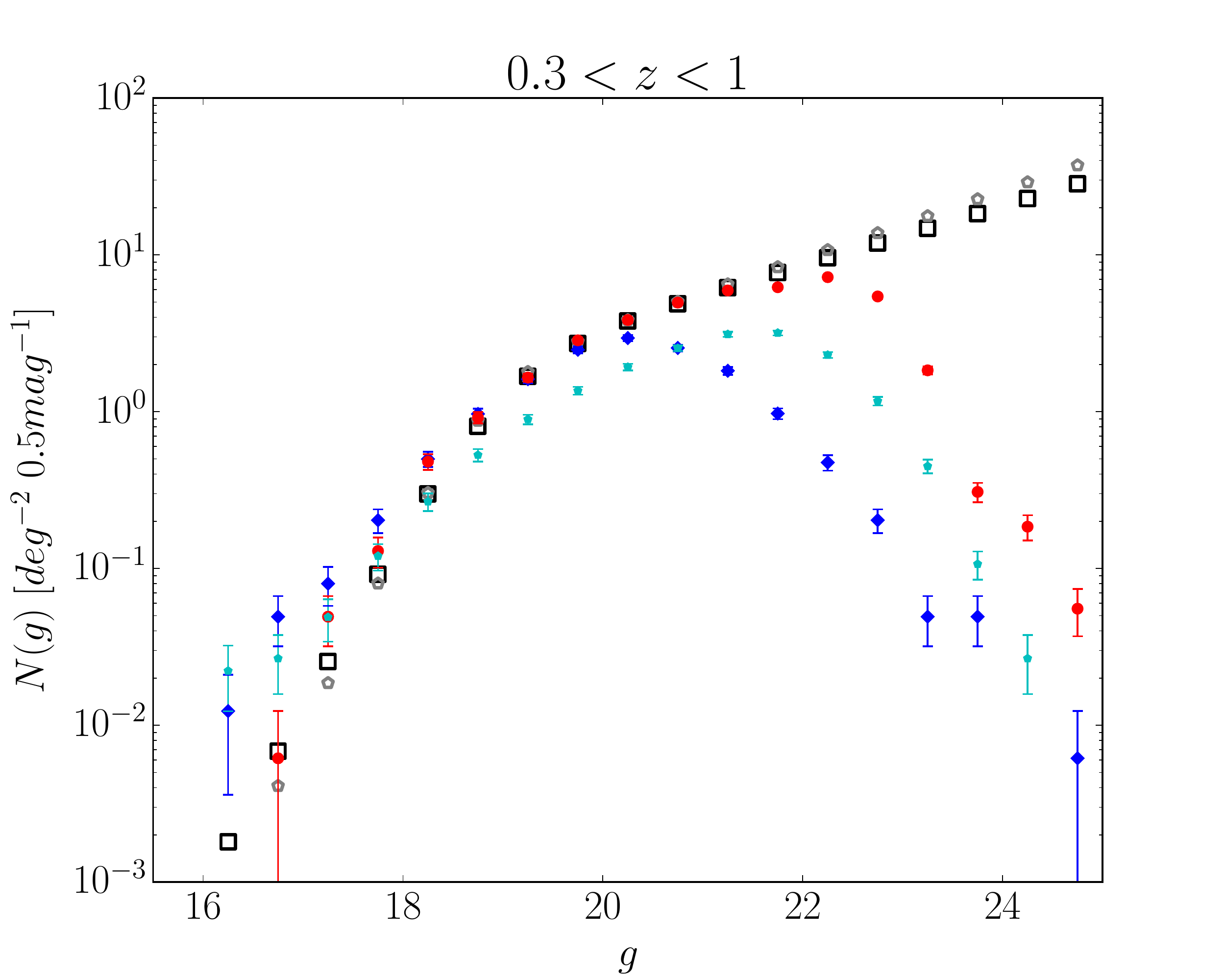}}
  \hspace{-0.7cm}
  \subfigure{
    \includegraphics[width=3.5in]{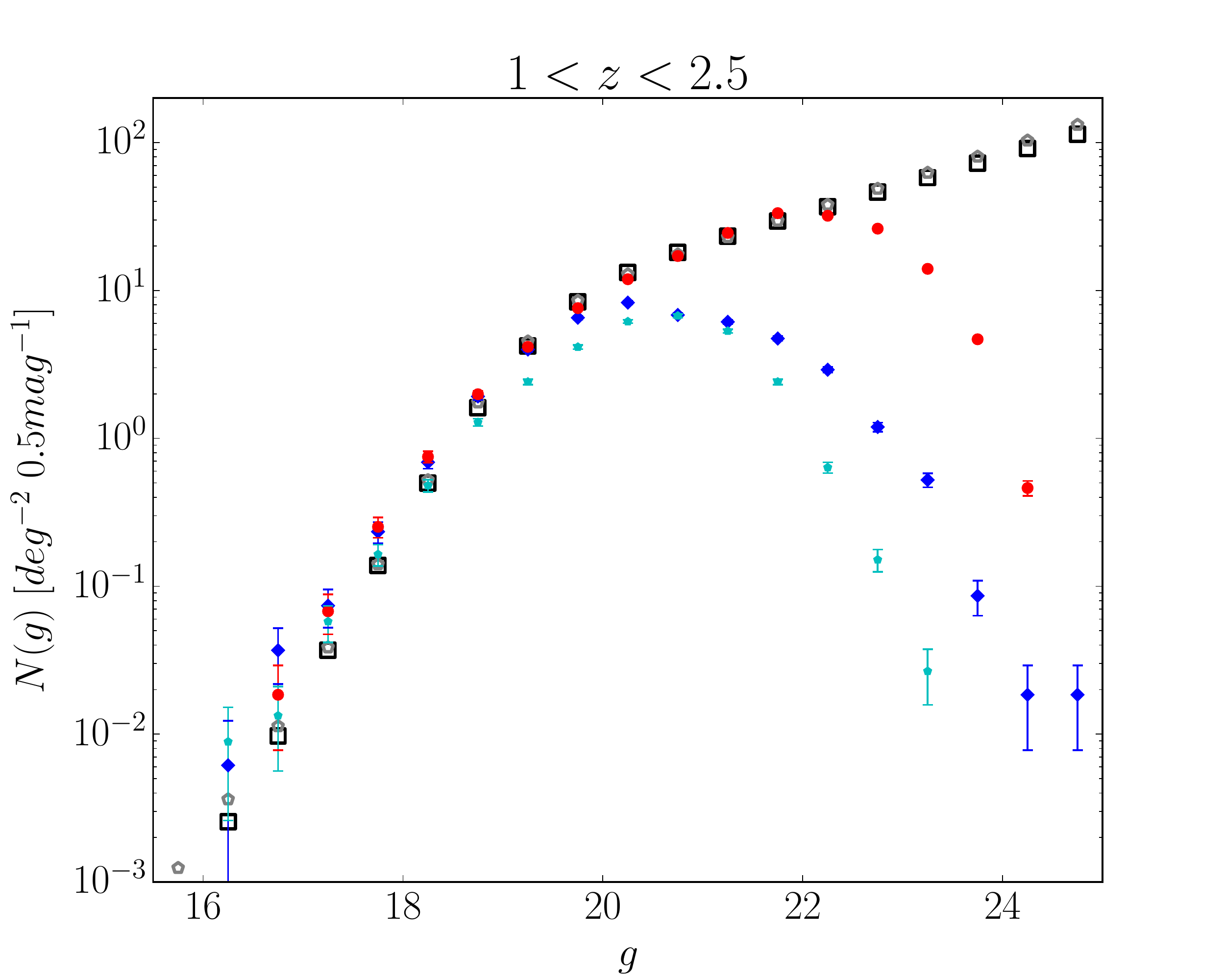}}
  \vspace{-0.5cm}
  \hspace{-0.5cm}
  \subfigure{
    \includegraphics[width=3.5in]{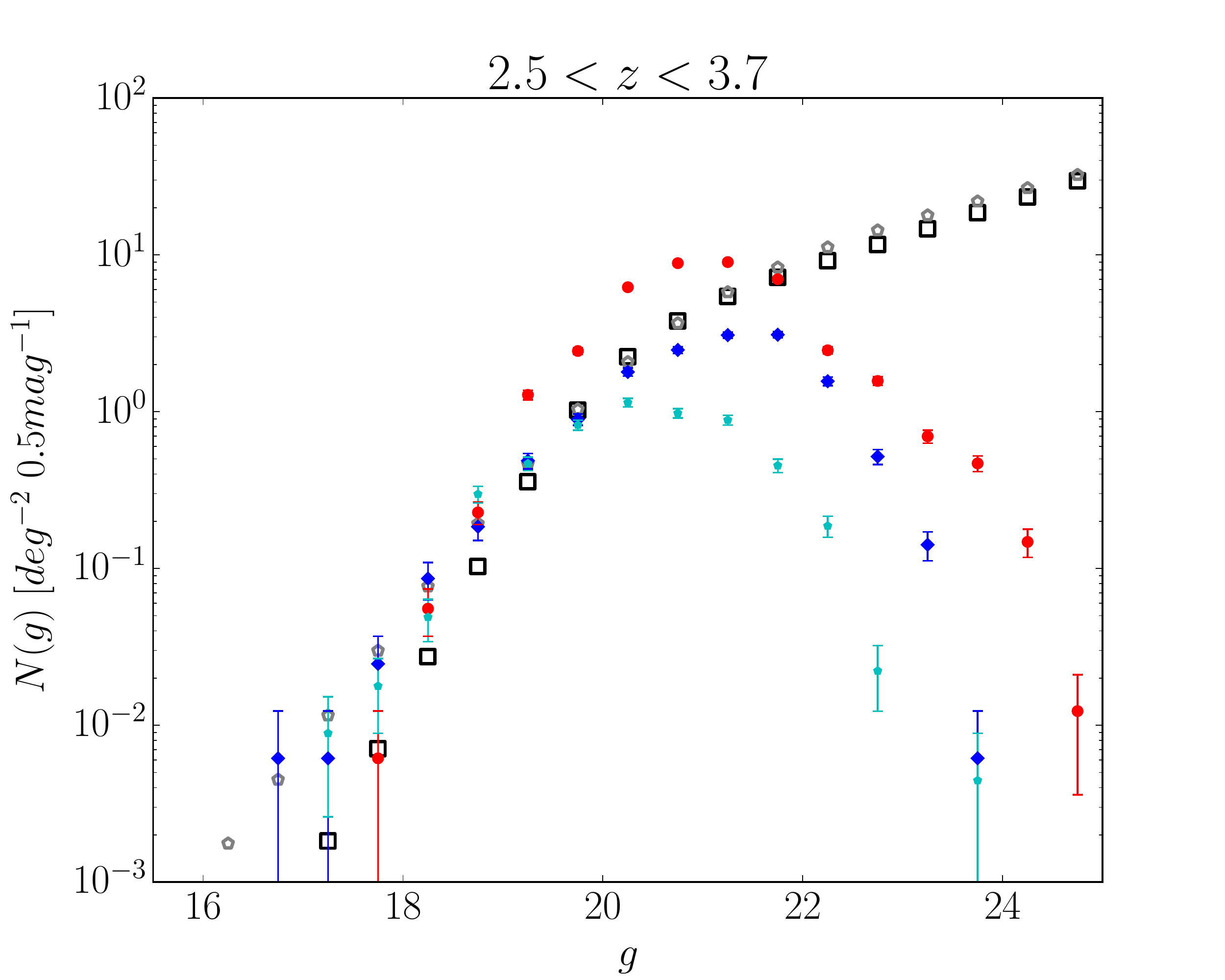}}
  \hspace{-0.7cm}
  \subfigure{
    \includegraphics[width=3.5in]{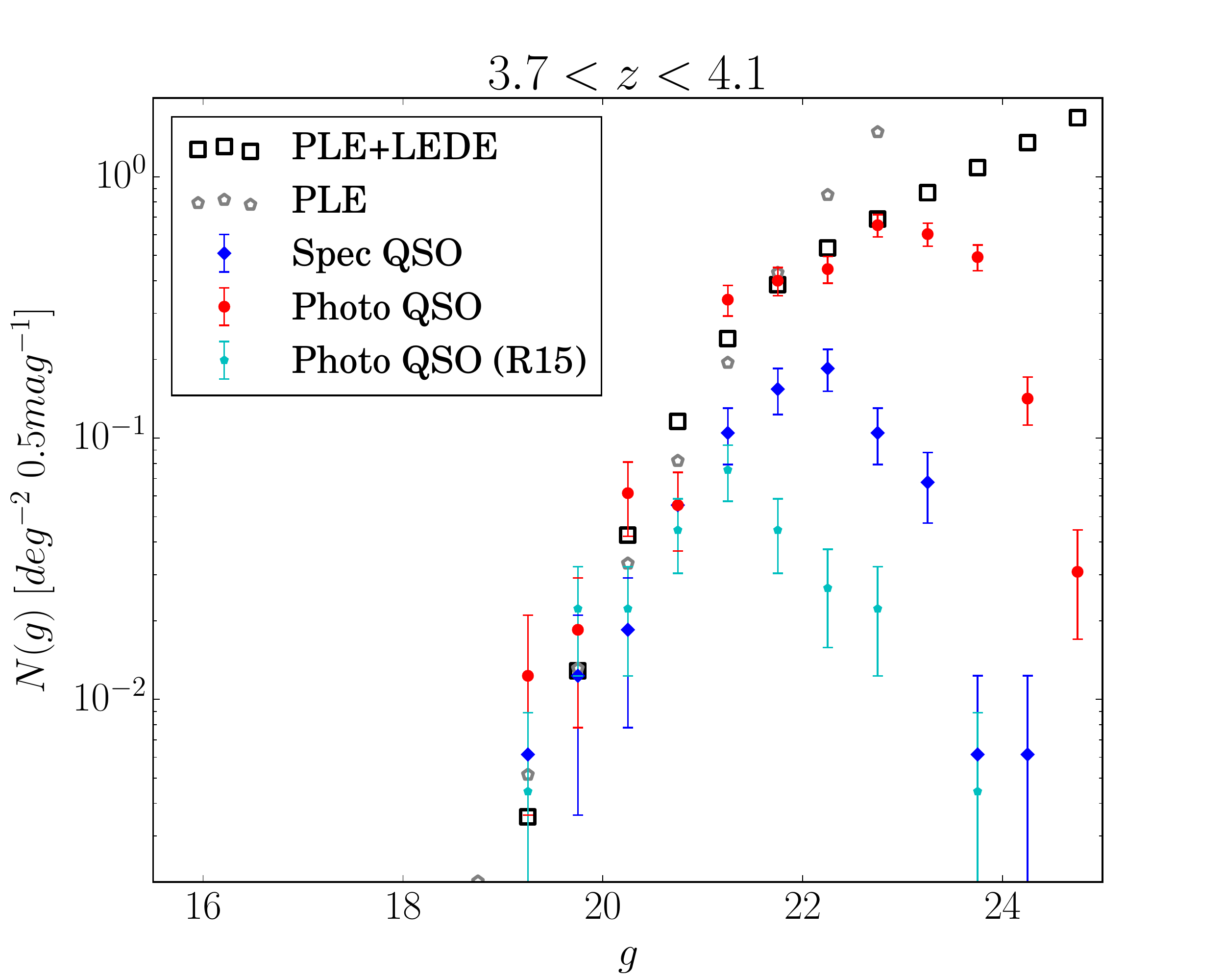}}
  \vspace{1cm}
  \caption{\label{fig:counts}The quasar number counts per deg$^2$ per 0.5mag in S82 ($340^{\circ}<{\rm R.A.}<45^{\circ}$, $-1.25^{\circ}<{\rm Decl.}<1.25^{\circ}$) as a function of $g$-band magnitude. The black open boxes and gray open pentagons are results calculated from the QLF \citep{NPD2016} PLE+LEDE model and PLE model, respectively. The blue diamonds represent the statistical results from spectroscopically identified quasars in this region, and the red dots show the results from our photo-$z$ QSO sample using DECaLS and WISE photometry. The photometric results of \citet{Richards2015} from the SDSS and WISE/Spitzer photometric data are also presented (cyan pentagons). The photo-$z$ QSO sample with DECaLS reaches a fainter magnitude limit than the spectroscopically identified samples. The photo-$z$ QSO sample is complete and efficient except for at $2.5<z<3.7$.}
\vspace{0.5cm}
\end{figure*}

\begin{figure}[htbp]
\hspace*{-1cm}
\epsscale{1.4}
\plotone{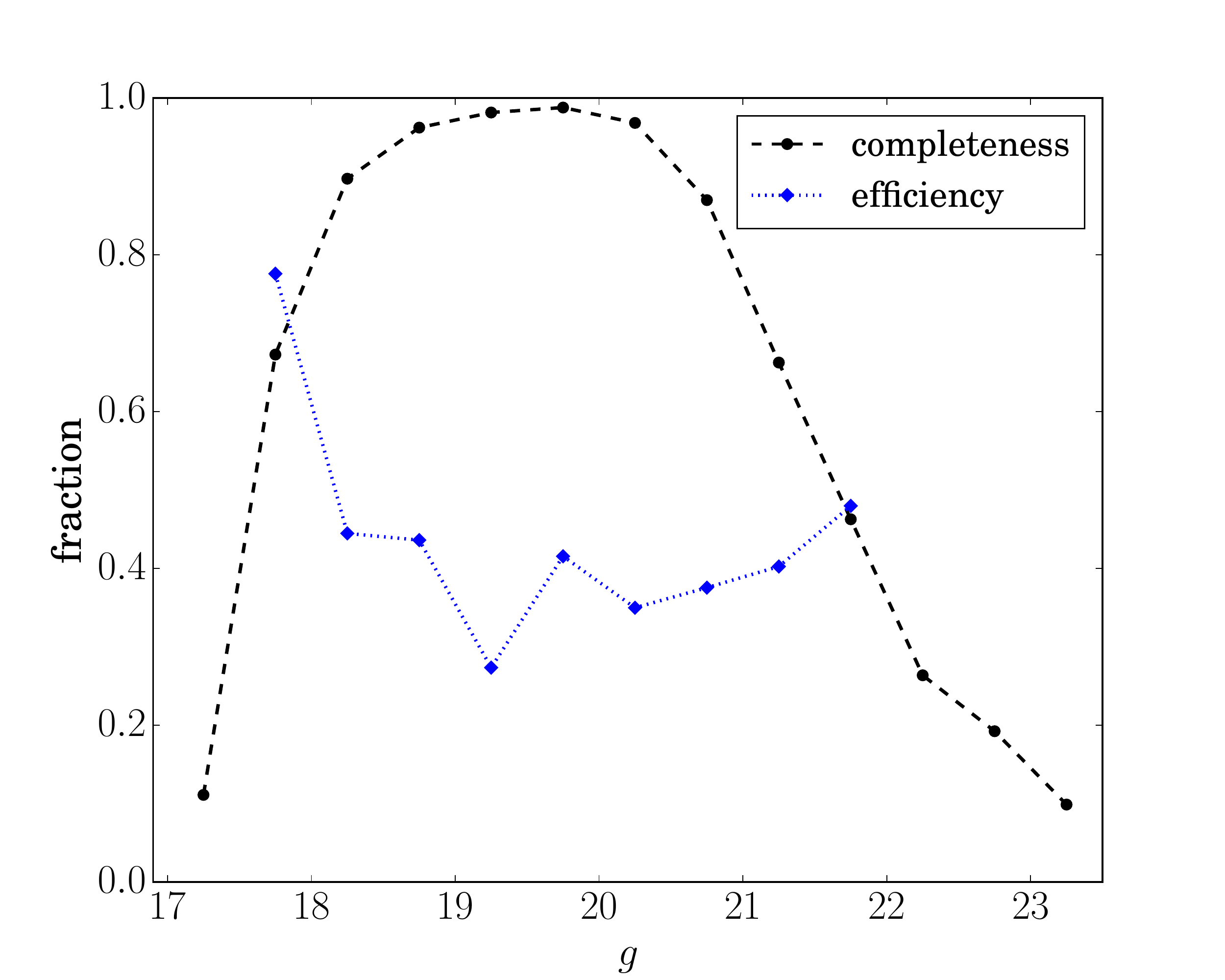}
\vspace{-0.5cm}
\caption{\label{fig:eff}The completeness (black dots) and efficiency (blue diamonds) as a function of $g$-band magnitude at $2.5<z<3.7$. The efficiency decreases to $\sim$ 40\% at $18<r<22$.}
\end{figure}

\subsection{Quasar Number Count Statistics in S82} \label{subsec:counts}
In a larger S82 region within $340^{\circ}<{\rm R.A.}<45^{\circ}$, $-1.25^{\circ}<{\rm Decl.}<1.25^{\circ}$ (roughly $b<-50^{\circ}$), there are 45,505 $r<23$ mag objects that satisfy the criteria of Equations (\ref{eq:30})-(\ref{eq:33}), listed in Table \ref{tab:catalog}. There are 12,332 spectroscopically identified quasars in this region, and the photometric quasar catalog covers 10,457 of them, with a completeness of 86\% at $r<22.5$ mag. \citet{Jiang2006} presented a spectroscopic sample of faint quasars, including 414 quasars down to $g = 22.5$ mag over $\sim 3.9$ deg$^2$. There are 282 quasars in their sample located within our catalog region, and 241 (85\%) of them are included in the photometric quasar catalog, and 32 are missed because of the morphology criterion.

Figure \ref{fig:dist} shows the $r$-band magnitude versus photo-$z$ distribution of the photometric quasars. There is a desert region at redshift $2.5<z<3.7$ and magnitude $r>21.5$. Because the colors of $2.5<z<3.7$ quasars are very close to those of stars, these quasars are excluded by criteria Equation \ref{eq:33} and \ref{eq:34}, shown in Figure \ref{fig:completeness}. The main contaminants show up at photo-$z\sim 2.7$ or photo-$z\sim 3.2$ and at $18.5<r<21.5$ (stars or galaxies). The selection completeness and efficiency at this redshift range is relatively low. There are some contaminant galaxies showing up at photo-$z<0.7$ and $20.5<r<22.5$. Slight contaminants show up at photo-$z\sim0.15$ and $r\sim 17.7$ (stars), photo-$z\sim 1$ (mainly stars), and photo-$z\sim 4.3$ (stars or galaxies).

Figure \ref{fig:counts} shows the number counts of the photometric quasar catalog as a function of $g$-band magnitude in redshift bins $0.3<z<1$ (top left panel), $1<z<2.5$ (top right panel), $2.5<z<3.7$ (bottom left panel) and $3.7<z<4.1$ (bottom right panel). The black open boxes and gray open pentagons are results calculated from the QLF \citep{NPD2016} PLE+LEDE model and PLE model, respectively. The results from our photometric quasar sample are shown as red dots, and the spectroscopic quasar number counts are shown as blue diamonds. \citet{Richards2015} presented a Bayesian quasar classification based on the SDSS optical data, and WISE/Spitzer mid-infrared photometry, and near-infrared data when available. We also plot the photometric quasar results (cyan pentagon) from \citet{Richards2015}. Because of the inclusion of deep DECaLS photometry, our photometric quasar sample reaches a fainter magnitude than the spectroscopic quasar sample in S82. In addition, we achieve a higher completeness even in the bright end compared to the photometric quasar sample in \citet{Richards2015} (R15). However, there are more $2.5<z<3.7$ photometric quasars showing up than the QLF. Figure \ref{fig:eff} shows the completeness and efficiency of quasar candidates in the range $2.5<z<3.7$. The efficiency decreases to $\sim$ 40\% at $18<r<22$. Only $g$, $r$, $z$, W1, and W2 are used in this case. More robust photometric quasar selection in this redshift range would need more photometric data in other bands. The selection completeness and efficiency at $3.7<z<4.1$ is high, so it is a redshift interval useful for spectroscopic surveying to study the QLF at redshift $z\sim4$.

\begin{table*}
\small
\begin{center}
\tablewidth{1pt}
\caption{Photometric quasar sample in S82} \label{tab:catalog}
\begin{tabular}{ccccrcccccc}
\tableline
\tableline
${\rm R.A.(J2000)}$ & ${\rm Decl.(J2000)}$ & ${\rm g}$ & ${\rm r}$ & ${\rm photo-}z$ & $z_1$ & $z_2$ & $P_{\rm prob}$ & $L_{\rm QSO}$ & $L_{\rm star}$ & $L'_{\rm Galaxy}$ \\
(deg) & (deg) & (mag) & (mag) & & & & & & & \\
\tableline
0.00193188&	-0.22927938&	20.23&	19.67&	3.45&	3.35&	3.60&	0.328&	-1.053&	-18.464&	-321.392\\
0.00370984&	-0.23459069&	23.28&	22.53&	1.45&	1.40&	1.55&	0.448&	-1.783&	-32.538&	-13.792\\
0.00677980& 0.58319355&	21.87&	21.81&	2.10&	1.85&	2.20&	0.548&	0.294&	-44.691&	-39.144\\
0.00741207&	0.30379374&	22.44&	22.33&	1.75&	1.70&	1.90&	0.546&	-1.892&	-50.650&	-19.015\\
0.00758513&	1.07482973&	23.47&	22.48&	1.45&	1.40&	1.50&	0.465&	-1.682&	-56.578&	-23.730\\
0.00783493&	-0.34638540&	21.93&	20.61&	0.30&	0.25&	0.35&	0.420&	-1.795&	-37.288&	-13.842\\
\tableline
\end{tabular}
\tablecomments{$r<23$ mag photometric quasars selected from DECaLS and WISE photometry with the criteria in Equations (\ref{eq:30})-(\ref{eq:34}) in the S82 region within $340^{\circ}<{\rm R.A.}<45^{\circ}$ and $-1.25^{\circ}<{\rm Decl.}<1.25^{\circ}$ (roughly $b<-50^{\circ}$). $z_1$ and $z_2$ are the lower and upper limits of ${\rm photo-}z$ in Equation \ref{eq:8}. (This table is available in its entirety in machine-readable form.)}
\end{center}
\end{table*}

\subsection{Classification using Random Forests}
An alternative method to decide the quasar criteria is machine learning classification procedure. We present a test using the Breiman and Cutler's Random Forests\footnote{https://www.stat.berkeley.edu/~breiman/RandomForests/} \citep{Breiman2001} classification with R $randomForest$ package\footnote{https://cran.r-project.org/web/packages/randomForest}. In this test, we use the same training data with that used in Section \ref{subsec:selection}, namely quasars and other point sources in a 15 $deg^2$ area in S82, and the same parameters, which are $L_{\rm QSO}$, $L_{\rm QSO} - L_{\rm Star}$, and $L_{\rm QSO}-L'_{\rm Galaxy}$. We use 100 trees, and it is sufficient for this classification case. Running a test on 98,450 quasars with DECaLS photometry, random forests method recovers 71\% of them. Only 19,853 objects are selected in the whole S82 region, and there are 0.3 stars per $deg^2$ and 0.4 galaxies per $deg^2$ at $r < 22.5$ with spectra in the SDSS DR13 catalog. Therefore, to select quasars, random forests classification method achieves higher efficiency but lower completeness than the selection criteria described in Section \ref{subsec:selection}.

\section{Summary} \label{sec:summary}
We present a new photo-$z$ regression algorithm for quasars considering the skew features of quasar color distributions, and use multivariate Skew-t funcitons to model the posterior relative flux distribution of quasars as a function of redshift and magnitude. The photo-$z$s are calculated by combining the posterior probability with the prior probability from the QLF. Photometric uncertainties are considered both in the photo-$z$ regression and classification procedures. The Skew-t photo-$z$ algorithm achieve a higher photo-$z$ accuracy than the XDQSOz and CZR method, and a higher calculation speed than the XDQSOz method. In the case that only the five SDSS bands are used, we achieve a photo-$z$ accuracy $R_{0.1}$ of 74\%. When combining SDSS/PS1/DECaLS optical photometry with WISE mid-infrared photometry, the photo-$z$ accuracy $R_{0.1}$ is enhanced to 87\%, 79\%, and 72\%, respectively. With WISE photometry, the degeneracy between $z\sim0.8$ and $z\sim 2.2$ is alleviated. The photo-$z$ accuracy decreases at $z<3.5$ due to the lack of $u$-band photometry when using PS1 or DECaLS photometric data.

To separate quasars from stars and galaxies, we perform a Milky Way synthetic simulation with the Besan\c{c}on model and galaxy template fitting. Quasars are selected with Bayesian probability criteria. We test the classification method based on the DECaLS optical and WISE mid-infrared data. The quasar selection completeness is higher than 70\% for a wide redshift range $0.5<z<4.5$, and a wide magnitude range $18<r<21.5$ mag. The photo-$z$ QSO sample with DECaLS reaches roughly mag fainter than the SDSS photo-$z$ QSO in R15. We find that the completeness at $z\sim2.8$ drops to $70\%$ if using only the $g$, $r$, $z$, W1 and W2 bands, because $z\sim 2.8$ quasars are close to the stellar locus. The completeness decreases at $z<1$, likely caused by quasar variability, morphology, and host galaxy. In a S82 test region with a high surface density of spectroscopically identified quasars, we recover 84\% of the quasars using our classification method. Meanwhile, only a small fraction of stars and galaxies with spectra in the SDSS DR13 are selected in this region. We also test the classification method in the VVDS deep field and COSMOS field. We recover 71\% and 76\% of the spectroscopically identified AGNs at $r<22.5$ mag. We present a catalog of 45,505 photometric quasars with $r<23$ mag in S82 using the DECaLS $grz$ and WISE W1 and W2 photometry. The sample is highly complete at $r < 22$ mag. The selection efficiency is high except for those with redshift at $2.5<z<3.7$. More photometric data in other bands are needed to improve quasar selection at $2.5<z<3.7$. 

Our photo-$z$ algorithm has a potential for the future LSST survey. To derive the QLF from the photometric quasar sample, careful correction for the selection completeness and efficiency is needed. A simulated quasar sample can be used to check the selection completeness \citep[e.g.,][]{Fan2001, Fan2003, Jiang2008, McGreer2013, Yang2016}. A simulated star sample and a galaxy sample can be used to check the effects of stars and galaxies on the quasar selection efficiency, respectively. Underestimation of the simulated star number counts, underestimation of the galaxy luminosity function, and overestimation of the galaxy size distribution will lead to overestimation of the selection efficiency, and overestimation of the QLF. Our quasar candidate selection method can be extended for multi-band photometric data, such as the optical photometric data from PS1; some future dataset, such as LSST data, Euclid \citep{Laureijs2011} data, and the Wide-Field Infrared Survey Telescope \citep[WFIRST,][]{Spergel2013} data; near-infrared data, such as the UKIRT Infrared Deep Sky Survey \citep[UKIDSS,][]{Lawrence2007}, the UKIRT Hemisphere Survey \citep[UHS,][]{Lawrence2013}, the VISTA Hemisphere Survey \citep[VHS,][]{McMahon2013} and the VISTA Kilo-degree INfrared Galaxy survey (Viking); and mid-infrared data WISE, NEOWISE \citep{Mainzer2011} and unWISE. More robust quasar candidate selection can be achieved by combining the probability classification with other methods, such as variability \citep[e.g.,][]{Geha2003, Schmidt2010, NPD2011} and proper motion \citep[e.g.,][]{Koo1986, Brunzendorf2001, Richards2009a}, radio surveys \citep[e.g.,][]{Becker2000, Lu2007}, and X-ray surveys \citep[e.g.,][]{Boyle1993, Trump2009}.\\

We gratefully acknowledge the support from the Ministry of Science and Technology of China under grant 2016YFA0400703, NSFC grants No.11373008 and 11533001, and the National Key Basic Research Program of China 2014CB845700.

We thank Yang Huang, Jo Bovy, Leo Girardi, Arjun Dey, and N. Palanque-Delabrouille for very helpful discussions. We thank David G. Grier for providing their KDE codes, Eduardo Ba{\~n}ados for help providing the PS1 photometric data, Chengpeng Zhang for providing their density contour plots code, and Michael Brown for providing the galaxy SEDs.

We acknowledge the use of SDSS photometric data. Funding for SDSS-III has been provided by the Alfred P. Sloan Foundation, the Participating Institutions, the National Science Foundation, and the U.S. Department of Energy Office of Science. The SDSS-III website is \url{http://www.sdss3.org/}. SDSS-III is managed by the Astrophysical Research Consortium for the Participating Institutions of the SDSS-III Collaboration including the University of Arizona, the Brazilian Participation Group, Brookhaven National Laboratory, Carnegie Mellon University, University of Florida, the French Participation Group, the German Participation Group, Harvard University, the Instituto de Astrofisica de Canarias, the Michigan State/Notre Dame/JINA Participation Group, Johns Hopkins University, Lawrence Berkeley National Laboratory, Max Planck Institute for Astrophysics, Max Planck Institute for Extraterrestrial Physics, New Mexico State University, New York University, Ohio State University, Pennsylvania State University, University of Portsmouth, Princeton University, the Spanish Participation Group, University of Tokyo, University of Utah, Vanderbilt University, University of Virginia, University of Washington, and Yale University.

We acknowledge the use of PS1 photometric data. The PS1 has been made possible through contributions by the Institute for Astronomy, the University of Hawaii, the Pan-STARRS Project Office, the Max-Planck Society and its participating institutes, the Max Planck Institute for Astronomy, Heidelberg and the Max Planck Institute for Extraterrestrial Physics, Garching, The Johns Hopkins University, Durham University, the University of Edinburgh, Queen's University Belfast, the Harvard-Smithsonian Center for Astrophysics, the Las Cumbres Observatory Global Telescope Network Incorporated, the National Central University of Taiwan, the Space Telescope Science Institute, the National Aeronautics and Space Administration under Grant No. NNX08AR22G issued through the Planetary Science Division of the NASA Science Mission Directorate, the National Science Foundation under Grant No. AST-1238877, the University of Maryland, and Eotvos Lorand University (ELTE).

We acknowledge the use of DECaLS photometric data, and the website is \url{http://legacysurvey.org}. This publication makes use of data products from the \emph{Wide-field Infrared Survey Explorer}, which is a joint project of the University of California, Los Angeles, and the Jet Propulsion Laboratory/California Institute of Technology, funded by the National Aeronautics and Space Administration. This work has made use of the TOPCAT \citep{Taylor2005}. We thank the Chinese Virtual Observatory, and the website is \url{http://www.china-vo.org}.

\software{%
$ADGofTest$    (\url{https://cran.r-project.org/package=ADGofTest}),
GALAXIA \citep{Sharma2011},
$kSamples$     (\url{https://cran.r-project.org/package=kSamples}),
$pracma$       (\url{https://cran.r-project.org/package=pracma}),
$randomForest$ (\url{https://cran.r-project.org/package=randomForest}),
$sn$           (\url{https://cran.r-project.org/package=sn}),
Skewt-QSO \citep{SkewtQSOv1},
XDQSOz \citep{Bovy2012}
}


\bibliography{ref}

\end{document}